\documentclass[article]{jss}\usepackage[]{graphicx}\usepackage[]{color}
\makeatletter
\def\maxwidth{ %
  \ifdim\Gin@nat@width>\linewidth
    \linewidth
  \else
    \Gin@nat@width
  \fi
}
\makeatother

\definecolor{fgcolor}{rgb}{0.345, 0.345, 0.345}
\definecolor{shcolor}{rgb}{0.94, 0.94, 0.94}
\newcommand{\hlnum}[1]{\textcolor[rgb]{0.686,0.059,0.569}{#1}}%
\newcommand{\hlstr}[1]{\textcolor[rgb]{0.192,0.494,0.8}{#1}}%
\newcommand{\hlcom}[1]{\textcolor[rgb]{0.678,0.584,0.686}{\textit{#1}}}%
\newcommand{\hlopt}[1]{\textcolor[rgb]{0,0,0}{#1}}%
\newcommand{\hlstd}[1]{\textcolor[rgb]{0.345,0.345,0.345}{#1}}%
\newcommand{\hlkwa}[1]{\textcolor[rgb]{0.161,0.373,0.58}{\textbf{#1}}}%
\newcommand{\hlkwb}[1]{\textcolor[rgb]{0.69,0.353,0.396}{#1}}%
\newcommand{\hlkwc}[1]{\textcolor[rgb]{0.333,0.667,0.333}{#1}}%
\newcommand{\hlkwd}[1]{\textcolor[rgb]{0.737,0.353,0.396}{\textbf{#1}}}%

\usepackage{framed}
\makeatletter
\newenvironment{kframe}{%
 \def\at@end@of@kframe{}%
 \ifinner\ifhmode%
  \def\at@end@of@kframe{\end{minipage}}%
  \begin{minipage}{\columnwidth}%
 \fi\fi%
 \def\FrameCommand##1{\hskip\@totalleftmargin \hskip-\fboxsep
 \colorbox{shadecolor}{##1}\hskip-\fboxsep
     \hskip-\linewidth \hskip-\@totalleftmargin \hskip\columnwidth}%
 \MakeFramed {\advance\hsize-\width
   \@totalleftmargin\z@ \linewidth\hsize
   \@setminipage}}%
 {\par\unskip\endMakeFramed%
 \at@end@of@kframe}
\makeatother

\definecolor{shadecolor}{rgb}{.97, .97, .97}
\definecolor{messagecolor}{rgb}{0, 0, 0}
\definecolor{warningcolor}{rgb}{1, 0, 1}
\definecolor{errorcolor}{rgb}{1, 0, 0}
\newenvironment{knitrout}{}{} % an empty environment to be redefined in TeX

   \let\hlslc\hlcom 
\usepackage{alltt}

\usepackage{thumbpdf,lmodern}
\usepackage{framed}

\usepackage[utf8]{inputenc}
\usepackage{amsmath,amsthm,amsfonts,amssymb,amscd}
\usepackage{lastpage}
\usepackage{enumerate}
\usepackage{mathrsfs}
\usepackage{graphicx}
\usepackage[nottoc]{tocbibind}
\usepackage[english]{babel}
\usepackage{float}
\usepackage{natbib}
\usepackage{comment}
\usepackage{xcolor}
\newcommand{\UA}{U\!A}
\newcommand{\UB}{U\!B}

\author{Edwin Farley\\Division of Applied Mathematics, Brown University
   \AND Roee Gutman\\Department of Biostatistics, Brown University}
\Plainauthor{Preston Schwartz\\Division of Applied Mathematics, Brown University}

\title{A Bayesian Approach to Linking Data Without Unique Identifiers}
\Plaintitle{Bayesian Data Linkage}
\Shorttitle{Bayesian Data Linkage}

\Abstract{
  Existing file linkage methods may produce sub-optimal results because they consider neither the interactions between different pairs of matched records nor relationships between variables that are exclusive to one of the files. In addition, many of the current methods fail to address the uncertainty in the linkage, which may result in overly precise estimates of relationships between variables that are exclusive to one of the files. Bayesian methods for record linkage can reduce the bias in the estimation of scientific relationships of interest and provide interval estimates that account for the uncertainty in the linkage; however, implementation of these methods can often be complex and computationally intensive. This article presents the \pkg{gfs\_sampler} package for the \proglang{Python} programming language that utilizes a Bayesian approach for file linkage. The linking procedure implemented in \pkg{gfs\_sampler} samples from the joint posterior distribution of model parameters and the linking permutations. The algorithm approaches file linkage as a missing data problem and generates multiple linked data sets. For computational efficiency, only the linkage permutations are stored and multiple analyses are performed using each of the permutations separately. This implementation reduces the computational complexity of the linking process and the expertise required of researchers analyzing linked data sets. We describe the algorithm implemented in the \pkg{gfs\_sampler} package and its statistical basis, and demonstrate its use on a sample data set.
}

\Plainkeywords{file linking, Bayesian analysis, multiple imputation, mcmc, R, Python}

\Address{
  Roee Gutman\\
  Department of Biostatistics\\
  Brown University 
  Box G-S121-7 
  121 S Main St, 7th Floor 
  Providence, RI 02912
  E-mail: \email{roee\_gutman@brown.edu}\\ }
\IfFileExists{upquote.sty}{\usepackage{upquote}}{}
\begin{document}

%% -- Introduction -------------------------------------------------------------

%% - In principle "as usual".
%% - But should typically have some discussion of both _software_ and _methods_.
%% - Use \proglang{}, \pkg{}, and \code{} markup throughout the manuscript.
%% - If such markup is in (sub)section titles, a plain text version has to be
%%   added as well.
%% - All software mentioned should be properly \cite-d.
%% - All abbreviations should be introduced.
%% - Unless the expansions of abbreviations are proper names (like "Journal
%%   of Statistical Software" above) they should be in sentence case (like
%%   "generalized linear models" below).

\newpage 

\section{Introduction} 

In health care and the social sciences, individual subjects' characteristics and outcomes are often dispersed over multiple files. To investigate relationships between variables that appear in different data sources, researchers seek to link individuals across these data sources while adapting to privacy regulations. In some record linkage applications, such as fraud detection and law enforcement, identifying records that belong to the same individual is essential; however, in many record linkage applications in epidemiology, medicine, and biostatistics, it is the preservation of associations between variables that is crucial, while the identification of individuals across data sets is not \citep{DOrazio}.

The statistical literature describes two broad classes of methods to link different data sources: statistical matching and record linkage. The objective of statistical matching algorithms is to learn about relationships among variables that are not jointly observed in a single data source \citep{Rassler,Rodgers1984}. The data sources may comprise a disjoint set of units, and the linking variables will typically be scientifically relevant variables, as opposed to identifying labels. Associations mediated through the linking variables can be estimated, but associations conditional on the linking variables cannot \citep{Rubin1974}. 

In record linkage applications, the linked files represent overlapping units such that records in different files represent similar entities \citep{FS,winkler2002methods}. Record linkage procedures can be classified into two main types of algorithms: \textit{deterministic} and \textit{probabilistic}. Deterministic record linkage methods identify records that belong to the same entity based on a deterministic agreement function applied to data elements that are common to both records. Probabilistic record linkage methods link records across data sets based on probabilities that pairs of records from the two different files represent the same unit. These probabilities are commonly estimated from the distribution of elements' agreement in the observed data or from a previously identified subset of records.

Deterministic methods are widely used and can be as simple as establishing that records represent the same units when they match exactly on one or more common data elements, such as first and last names. Deterministic linking based on perfect agreements has been shown to have a higher rate of true links than probabilistic linking; however, when the underlying data elements are subject to error in recording, deterministic linking may have a high level of missed true links compared to probabilistic linking \citep{gomatam2002empirical,campbell2008record}.

We describe a Bayesian probabilistic linkage algorithm that views record linkage as a missing data problem. The algorithm accounts for relationships between pairs of records and relationships between variables exclusive to one of the files. The algorithm is implemented in the \pkg{gfs\_sampler} package for the \proglang{Python} programming language \citep{python}. The paper proceeds as follows: Section \ref{sc:bg} provides a brief review of existing record linkage procedures and their implementations. Section \ref{sc:meth} describes the proposed Bayesian record linking procedure. Section \ref{sc:sampling} describes the Markov chain Monte Carlo (MCMC) sampling algorithm for the proposed Bayesian record linkage procedure. Section \ref{sc:example} provides an empirical example demonstrating the use of \pkg{gfs\_sampler}. Section \ref{sc:conclusion} provides a summary and a discussion of future work.

\newpage

\section{Probabilistic linkage and related work}
\label{sc:bg}

Probabilistic record linkage relies on a framework proposed by \citet{FS}. Let $\gamma_{ij} = (\gamma_{ij1},\ldots,\gamma_{ijP})$ be an agreement vector between record $i$ in file $A$ and record $j$ in file $B$, where $P$ is the number of covariates that appear in both files and $\gamma_{ijk} \in\{1,..,L_{k}\}$ represents the level of agreement of the values of covariate $k$ in the two files. The Fellegi and Sunter model assumes that $\gamma_{ij}$ follows a mixture distribution, such that if $(i,j)$ is a true link, $\gamma_{ij} \sim f_{M}(\gamma_{ij}|\theta_M)$, and if $(i,j)$ is a non-link, $\gamma_{ij} \sim f_{U}(\gamma_{ij}|\theta_U)$. Formally, the mixture model is:
\begin{equation}
\label{eq:fsMixture}
\Prob(\gamma_{ij}) = \pi f_{M}(\gamma_{ij}|\theta_M) + (1-\pi)f_{U}(\gamma_{ij}|\theta_U),
\end{equation}

where $\pi$ is the probability that a pair of records is a true link. 

Because in many applications $\pi,\theta_{M}$ and $\theta_{U}$ are unknown, the Expectation Maximization (EM) algorithm is commonly used to estimate these parameters \citep{belin1995method,larsen2001iterative}. Let $\hat{\pi},\hat{\theta}_{M}$ and $\hat{\theta}_{U}$ be the estimates of $\pi,\theta_{M}$ and $\theta_{U}$, respectively. The Fellegi and Sunter algorithm calculates weights for each pair of records $(i,j)$, 
%$$w_{ij} = \frac{f_{M}(\mathbf{\gamma_{ij}}|\hat{\theta}_M)}{f_{U}(\mathbf{\gamma_{ij}}|\hat{\theta}_U)}$$
$$w_{ij} = \frac{f_{M}(\gamma_{ij}|\hat{\theta}_M)}{f_{U}(\gamma_{ij}|\hat{\theta}_U)}.$$

These weights commonly inform a ``greedy'' algorithm that iteratively links and removes from the matching pool the pair of records with the highest probability of a match until a certain predefined probability threshold cannot be met by any remaining record pairs \citep{FS}. These remaining record pairs with weights that are below the threshold are either clerically reviewed or declared non-links. Although the computational simplicity of the algorithm is considered a strength, it may produce a globally sub-optimal linkage, because it does not consider the relationships between distinct pairs of records. One possible solution to this issue is to rely on an optimization algorithm that forces one-to-one linkage after estimation of $w_{ij}$ \citep{Jaro}. 

Estimation of the probabilities that pairs of records represent the same entities can be computationally intensive. ``Blocking'' is a file linkage technique that reduces the number of possible links. Once the data sets are ``blocked,'' only pairs of records that agree on the blocking variables are considered for linking \citep{NewcombeKennedy,newcombe1988handbook}. This reduces computational complexity and increases the accuracy of the linkage at the same time. \citet{murray2016probabilistic} described other possible indexing methods, and possible limitations of linkage error propagation when blocking is used. 

Despite the popularity of the Fellegi and Sunter algorithm for linking records, it has a number of weaknesses. First, the mixture model will always identify two clusters regardless of whether the two files include records that represent the same entities \citep{winkler2002methods}. Moreover, the Fellegi and Sunter algorithm assumes that linkage identification of different record pairs are independent \citep{sadinle2017bayesian}, and thus may not be efficient in identifying accurate links even with the inclusion of the optimization step proposed by Jaro \citep{Jaro}. Lastly, in many applications, the linkage process is not the final goal of the analysis, and adjustments for errors in the linkage process should be addressed when estimating relationships between variables that are exclusive to one of the files. To address this issue, estimation procedures that rely on the non-informative linkage assumption have been proposed \citep{Lahiri,chambers2009}. These procedures also assume that the probabilities that a pair $(i,j)$ is a true link, $\Prob((i,j)\in M|\mathbf{\gamma_{ij}})$, or a false link, $\Prob((i,j)\in U|\mathbf{\gamma_{ij}})$, are known or can be estimated accurately. The non-informative linking assumption asserts that associations between variables that are exclusive to one of the files do not inform the linkage. Based on this assumption, $\Prob((i,j)\in M|\mathbf{\gamma_{ij}})$ and $\Prob((i,j)\in U|\mathbf{\gamma_{ij}})$ can be used as weights in a regression model \citep{Lahiri,han2019statistical}, or in generalized estimating equations \citep{chambers2009regression}. In many applications, the non-informative linkage assumption is violated  \citep{Gutman,GutmanSammartino}, which may lead to biased estimates. This violation is especially pronounced when the true links and the false links cannot be separated well using variables that appear in both files.

Bayesian procedures can address these shortcomings by introducing a latent linking structure that maps records in one file to records in another file \citep{Gutman,sadinle2017bayesian,steorts2016bayesian, dalzell2018regression}. These Bayesian models sample from the joint distribution of the latent linkage structure and model parameters. This sampling enables the future adjustments for error in the linking structure; however, sampling from this joint distribution can be computationally complex, especially when applied researchers are interested in examining multiple relationships between variables exclusive to one of the files. A possible solution is to treat the linkage structure as missing data, and use multiple imputation to propagate the linkage error to estimate these relationships \citep{rubin2004multiple}. This approach demands less time and expertise on behalf of researchers and allows for flexibility in analysis of the linked data set. In this approach, a number of linked data sets are generated using samples from the posterior distribution of the linkage structure. These data sets can be analyzed separately and final results are obtained using standard multiple imputation rules \citep{rubin2004multiple,barnard1999miscellanea}. In addition, some Bayesian linkage algorithms relax the non-informative linkage assumption by modeling the associations between variables that are exclusive to one file \citep{Gutman,dalzell2018regression}. 
 
%One such function matches a record in one data set to the record in the other data set that is closest in terms of its value in the empirical cumulative density function of a specified reference variable. A separate function takes from predictive mean matching allows for the specification of a regression model that is estimated over the same set of matching variables in both data sets and a matching decision is made based on the distance between the true and estimated value for the response variable for both estimated models \cite{StatMatch, DOrazio}. A hot deck method is used to propose matches, which could introduce bias in the resulting linkage \cite{Shlomo2019}.

Existing software packages for record linkage are based on the Fellegi and Sunter algorithm and have similar limitations. One \proglang{R} package is \pkg{fastLink}, which links records based on the mixture model in Equation \eqref{eq:fsMixture} \citep{enamorado_fifield_imai_2019}. The \pkg{fastLink} package estimates the parameters of the model using the EM algorithm, relying on efficient data structures to increase the speed of the estimation algorithm. To adjust for linkage error, it provides the probabilities that two records are a match, and these can be used as weights in regression models. An additional limitation of this package is that support for blocking is not well-integrated into the package, so its use may increase computation time. A different \proglang{R} package is \pkg{RecordLinkage} \citep{de_bruin_j_2019_3559043}. This package is slower than \pkg{fastLink} in terms of estimating the model parameters, but it can handle blocking in a simpler manner. Another \proglang{R} package is \pkg{StatMatch}. This package performs statistical matching, relying on predictive mean matching to impute variables that are not jointly observed in a single data source \citep{StatMatch}. This method implements the statistical matching procedure proposed by \citet{little1988missing}, and it does not assume that both files include similar entities, unlike the other two packages.

A principal reason that Bayesian record linkage is under utilized in applied research is the lack of statistical software that implements it. We describe a software package called \pkg{gfs\_sampler} that implements the Bayesian linkage algorithm described by \citet{Gutman}. The package is implemented in \proglang{Python}, but a set or wrapper functions for \proglang{R} is also available, with more details provided in the Appendix (\ref{sc:appendix}).

\section{The proposed method and notation}
\label{sc:meth}
%Some of the variables in $\mathbf{X}_{A,i}$ and $\mathbf{X}_{B,i}$ may represent the same attribute, but they may also represent different attributes. 
%\Edwin{The notation here is getting confusing, because we want to index X by file (A and B), by row, by column, and later by block. Maybe we should say that the hypothetical complete data set X is composed of (A, B, Z). We refer to Z on its own, so maybe we should start using A and B instead of subscripting X by A and B. That should give us cleaner notation for rows, etc.}
We assume that data on similar units is dispersed across two files, $A$ and $B$, comprised of $n_A$ and $n_B$ records, respectively. Let $\mathbf{X}_i = (X_{i,1},\ldots,X_{i,P})$, $i=1,\ldots,n_A$, denote a vector of covariates for entity $i$ in complete data set $\mathbf{X}$. The $P$ covariates are partitioned into three components of size $P_1$, $P_2$, and $P_3$, such that $P_1+P_2+P_3=P$. The first component includes $P_1$ covariates that are exclusive to file $A$, $\mathbf{XA}_{i} = (X_{i,1},\ldots,X_{i,P_{1}})$. The second component includes $P_2$ covariates exclusive to file $B$, $\mathbf{XB}_{i} =  (X_{i,P_1+1},\ldots,X_{i,P_1+P_2})$. The third component includes $P_3$ covariates that are recorded without error in both files. Let $\mathbf{Z}_i = (X_{i,P_1+P_2+1},\ldots,X_{i,P})$ denote the $P_3$ covariates in both files. The variables in $\mathbf{Z}$ can be used to create blocks and restrict the number of possible record pairs that are considered as links. 

Sampling from the posterior distribution of a linkage structure is computationally complex, and blocking is a practical technique to reduce computational complexity. We assume that records are partitioned into $J$ blocks defined by the values of $\mathbf{Z}$. In file $A$, each block $j \in \{1,\ldots,J\}$ comprises $I_{A_j}$ records. Similarly, in file $B$, block $j$ comprises $I_{B_j}$ records. We denote records from file $A$ in block $j$ as $\mathbf{XA}^j_{i}$, and records from file $B$ in block $j$ as $\mathbf{XB}^j_{i}$.

We introduce a latent structure $\mathbf{C}=\{\mathbf{C}_j : j = 1,\ldots,J \}$, where $\mathbf{C}_j$ represents the matching permutations in block $j$. For each record in file $A$, this structure indicates the matching record in file $B$, such that $\mathbf{C}_j \in \{C_{jk}: k = 1,...,K_j\}$, where $C_{jk}$ is one possible permutation for block $j$ and 

$$K_j = \frac{\left(\max\left(I_{A_j} ,I_{B_j}\right)\right)!}{ | I_{ A_j } - I_{B_j} | !} \enspace .$$

In the $j$th block, $C_{jk}[i] \in \{1,\ldots,I_{B_j}\}$ is the index for the record in file $B$ that is matched to record $i \in \{ 1, \ldots,  I_{A_j}\}$ in file $A$ according to permutation $k$. For a given matching permutation $k$ for block $j$, the linked data for record $i$ of file $A$ is $(\mathbf{XA}^j_{i},\mathbf{XB}^j_{C_{jk}[i]},\mathbf{Z}_i)$. For example, $C_{j5}[2] = 4$ implies that in the fifth possible linking permutation for block $j$, the second record in file $A$ is linked to the fourth record in file $B$. The newly-linked record is $(\mathbf{XA}^j_{2}, \mathbf{XB}^j_{4}, \mathbf{Z}_2)$. 

When $I_{A_j}=I_{B_j}$, all records in block $j$ of files $A$ and $B$ are linked. When $I_{A_j}\neq{B_j}$ some records in either file $A$ or file $B$ are left unmatched. Let $\UA_{j}$ be the set of indices of unmatched records in block $j$ of file $A$, and $\UB_{j}$ be the set of indices of unmatched records in block $j$ of file $B$. Given $\mathbf{C}_j = C_{jk}$ the density for one entity is:
\begin{eqnarray}\label{eq:cases}
L_{i,j}(\mathbf{\theta},\mathbf{C}_j = C_{jk}) & = & \begin{cases}
 f_{AB}(\mathbf{XA}^j_{i},\mathbf{XB}^j_{\unboldmath {C_{jk}}(i)}\mid\theta, \mathbf{Z}_{i})f_{Z}(\mathbf{Z}_{i}\mid\theta), & i\not\in \UA_j \\
 f_{A}(\mathbf{XA}^j_{i}\mid \mathbf{Z}_{i},\theta)f_{Z}(\mathbf{Z}_i\mid\theta) & i \in \UA_{j}  
\end{cases} \\[1ex]
L_{l,j}(\mathbf{\theta},\mathbf{C}_j = C_{jk}) & = &  
f_{B}(\mathbf{XB}^j_{l}\mid \mathbf{Z}_{i},\theta)f_{Z}(\mathbf{Z}_{l}\mid\theta) \hspace{6.4em} \ \ l \in \UB_j \nonumber
\end{eqnarray}
where $\theta$ is a parameter vector, $f_A$, $f_B$, and $f_{AB}$ are, respectively, the marginal densities of $\mathbf{XA}^j_{i}$ and $\mathbf{XB}^j_{l}$, and their joint density conditional on $\mathbf{Z}_i$, and $f_Z$ is the marginal density of $\mathbf{Z}_i$. Multiplying over cases in a block, the likelihood for $\theta$ and $\mathbf{C}_j$ for block $j$ is 

\begin{equation}\begin{split}\label{eq:likeblock}
L_j(\theta,\mathbf{C}_j) =  & f_{Z}(\mathbf{Z}_i\mid\theta)^{\max(I_{A_j},I_{B_j})} \times
   \left(\prod\limits_{i\in \UA_{j}}  f_{A}(\mathbf{XA}^j_{i}\mid \theta,\mathbf{Z}_i) \right) \times \\
  & \left(\prod\limits_{l\in \UB_{j}} f_{B}(\mathbf{XB}^j_{l}\mid \theta,\mathbf{Z}_l)\right) \times 
  \left(\prod \limits_{i\not\in \UA_{j}} 
    f_{AB}(\mathbf{XA}^j_{i},\mathbf{XB}^j_{C_{jk}(i)}\mid\theta,\mathbf{Z}_i)\right) 
\end{split}\end{equation}

The form of $f_{AB}$ is specific to an application, but it is often convenient to express it as a product of conditional distributions, for example
\begin{equation}\label{eq:conditional}\begin{split}
f_{AB}(\mathbf{XA}_{i},\mathbf{XB}_{i}\mid\theta,\mathbf{Z}_i) = & f_{A}(\mathbf{XA}_{i}\mid \mathbf{Z}_i,\theta_{A})\cdot f_{B|A}(\mathbf{XB}_{i}\mid \mathbf{Z}_i,\mathbf{XA}_{i},\theta_{B\cdot A}),
\end{split}
\end{equation} 
where $\theta = (\theta_A, \theta_{B \cdot A})$.
The densities $f_{A}$ and $f_{B|A}$ may represent models of scientific interest as well as models for relationships that are useful purely for identifying true links. An example of a model that is not of scientific interest may describe relationships between zipcode digits that appear in both files but are recorded inconsistently across files. 

A possible formulation of $f_{B|A}$ is based on a combination of univariate generalized linear models. Formally, 

\begin{equation}\label{eq:breakB1}
\begin{split}
f_{B|A}(\mathbf{XB}_{i}\mid \mathbf{Z}_i,\mathbf{XA}_{i},\theta_{B \cdot A}) = &  f_{B_1}(XB_{i,1}\mid \mathbf{Z}_i,\mathbf{XA}_{i},\theta_{1}) \times \\ & f_{B_2}(XB_{i,2}\mid \mathbf{Z}_i,\mathbf{XA}_{i},\mathbf{XB}_{i(-2)},\theta_{2}) \times \cdots \times \\ & f_{B_{P2}}(XB_{i,P_{2}} \mid \mathbf{Z}_i,\mathbf{XA}_{i},\mathbf{XB}_{i(-P_2)},\theta_{P_2}),
\end{split}
\end{equation}

where $\mathbf{XB}_{i(-p)} = (XB_{i,1},\ldots,XB_{i,p-1})$ and $\theta_{B\cdot A} = (\theta_{1},\ldots,\theta_{P2})$. In the \pkg{gfs\_sampler} package, depending on the data type of $XB_{i,p}$, each univariate conditional density in Equation \eqref{eq:breakB1} can be modeled as either a Normal linear regression model, a Poisson regression model with log link model, or a logistic regression model.

Equation \eqref{eq:cases} shows that we need the marginal distributions $f_{A}$ and $f_{B}$ as well as the joint distribution $f_{AB}$, which is defined in Equation \eqref{eq:conditional}. In blocks where $I_{A_j} \leq I_{B_j}$, deriving the marginal distribution $f_{B}$ from Equation \eqref{eq:conditional} requires integration over the $\mathbf{XA}_{i}$, which is often analytically intractable. To overcome this issue, we impute $I_{B_j} - I_{A_j}$ unobserved $\mathbf{XA}^j_i$ values in such blocks. This results in blocks with an identical number of records in both files, eliminating the need to define $f_{B}$ analytically. The unobserved $\mathbf{XA}^j_i$ are imputed by sampling from a uniform distribution over the observed records in block $j$ of file $A$. This distribution assumes that the unobserved records in $A$ follow the same distribution as the observed records in that block. In blocks where $I_{A_j} > I_{B_j}$, we choose only $I_{B_j}$ observations from file $A$ to be linked, exploiting the monotone missing data pattern \citep{little2019statistical}. 

To complete the Bayesian model we postulate a uniform independent prior distribution over the possible permutations $\mathbf{C}_j$, which are independent of $\theta$. In addition, $\theta_1,\ldots,\theta_{P2}$ are assumed to independently follow the default prior distributions in the \pkg{pymc3} package for \proglang{Python} \citep{Salvatier2016}. This results in a posterior distribution of
\begin{equation}\label{eq:postDist}
\Prob(\theta, \mathbf{C}| \mathbf{XA},\mathbf{XB},\mathbf{Z}) \propto \pi(\theta) \prod_{j=1}^{J}\sum_{k=1}^{K_j}L_j(\theta,C_{jk}),
\end{equation}
where $\pi(\theta)=\pi(\theta_A)\times\pi(\theta_{1})\times\cdots\times\pi(\theta_{P2})$ is the prior distribution of $\theta$.
%We are assuming that for all blocks $I_{B_j} \geq I_Aj$, the numerator for $K_j = I_{B_j} !$ for all $j$. All of the records in file $B$ that are not matched to records in file $A$ are designated a non-matched value. from the  in the file with the smaller block have a match in the file with the larger block, and for the missing rows, the procedure initially adds rows of NaNs to match the larger block. When there are fewer rows in $A$ than in $B$ for a certain block we can fill in the missing row with values sampled from the other rows in the block for the purposes of computing the regression coefficients.

%Note that there is no explicit ordering of permutations, the parameter $k$ is merely used for enumerating the choices.
%Furthermore, linkages are only sampled at the level of the blocks, and a full permutation of file $B$ is constructed by combining sampled permutations from within blocks.

%Extending our notation to cover the case $I_{A_j} \not= I_{B_j}$, we let $U_{A_j} = U_{A(C_j)$ be the set of indices of unmatched records in file A and $U_Bj = U_B(Cj)$ the corresponding set for file B.

\section{The three-step iterative sampling scheme} 
\label{sc:sampling}

To sample from the joint posterior distribution of the parameters $\theta$ and linking structure $\mathbf{C}$, the \pkg{gfs\_sampler} package relies on a Gibbs Sampler with a Metropolis-Hastings step within each block \citep{gelman2013bayesian}. The starting values for all $\mathbf{C}_j$ are determined based on the initial order of records in the given files. Let $\mathbf{C}^{(t)}$ be the resultant permutation at the completion of iteration $t$ and $\theta^{(t)}$ the values of the regression parameters at the completion of iteration $t$. Iteration $t \in 1,\ldots,T$ of the Gibbs sampling algorithm is:

\begin{itemize}
    \item[1.] Given $\mathbf{C}^{(t-1)}$, sample the regression parameter $\theta^{(t)}$ from Equation \ref{eq:postDist}. Because $\theta_{1},\ldots,\theta_{P2}$ are independent given $\mathbf{C}^{(t-1)}$, each sub-model in Equation \ref{eq:breakB1} is sampled separately using the No-U-Turn Sampler implemented in \pkg{pymc3} \citep{hoffman2014no}. 
    \item[2.] Given $\theta^{(t)}$, sample from $C_{jk}^{(t)}$ independently using the Metropolis-Hastings algorithm described by \citet{yingnian-95} and \citet{Gutman} for each block $j$ to obtain $\mathbf{C}^{(t)}$. 
    \item[3.] For each block $j$, if $I_{B_{j}} > I_{A_{j}}$, sample $I_{B_{j}} - I_{A_{j}}$ records from the $I_{A_{j}}$ observed records from block $j$ in file $A$ to link with the remaining records from file $B$ in block $j$.
\end{itemize}

Below we expand on the three steps of the Gibbs sampling procedure for block $j$.

\subsection*{Step 1: Regression sampling}
We assume independent prior distributions for each conditional model in Equation \ref{eq:breakB1}, so the parameters of each generalized linear regression model, $\theta_{1},\ldots,\theta_{P2}$, can be sampled independently from their respective posterior distributions. For example, let $\mathbf{XB}_{i} = (XB_{i,1}, XB_{i,2})$, such that $XB_{i,1}$ is continuous and $XB_{i,2}$ is binary. The following regression models would be used:
\begin{eqnarray}
\label{eq:modelExam}
    f_{B_1} :& XB_{{C}_{j}[i],1}|\mathbf{XA}_{i},\theta_1 \sim N(\mathbf{XA}_{i}\cdot\beta_1,\sigma^{2}) \nonumber\\ 
    f_{B_2} :& XB_{{C}_{j}[i],2}|XB_{{C}_{j}[i],1}, \mathbf{XA}_{i},\theta_2 \sim  Bernoulli(\text{logit}^{-1}((XB_{{C}_{j}[i],1},\mathbf{XA}_{i})\cdot\theta_2)),
\end{eqnarray}
%$$
%X_{Bi1}|\mathbf{X}_{Ai},\theta_1 \sim N(\mathbf{X}_{Ai}^{t}\beta_1,\sigma^{2}) \\
%X_{Bi2}|X_{Bi1},\mathbf{X_Ai},\theta_2 \sim %Bernoulli(logit^{-1}((X_{Bi1},\mathbf{X}_{Ai})^{t}\theta_2)), 
%$$
where $N$ is the Normal distribution, $\theta_1 = (\beta_1,\sigma^2)$, and $\text{logit}^{-1}(p) = \exp(p)/(1-\exp(p))$. We will assume independent prior distributions for each regression parameter, $\pi(\theta_{2},\theta_{1}) \propto \pi_{2}(\theta_{2})\pi_{1}(\theta_{1})$. Formally, 
\begin{eqnarray}
\label{eq:modelPrior}
    \pi_{1} :& \beta_{1}|\sigma \sim N(0,10^3 \times I_{P_1}), \quad \Prob(\sigma)  \propto 1    \nonumber\\ 
    \pi_{2} :& \theta_{2} \sim N(0,10^3\times I_{P_1 +1}),
\end{eqnarray}
where $I_{p}$ is a $p\times p$ identity matrix. Based on this prior distributions and likelihoods, the posterior distributions for $\theta_1$ and $\theta_2$ are independent given $\mathbf{C}_j$, and can be sampled independently from their conditional posterior distributions at each iteration of the Gibbs sampler:
\begin{eqnarray}
\Prob(\theta_1|\mathbf{XA},\mathbf{XB},\mathbf{Z},\mathbf{C}_j^{(t)}) \propto f_{B_1}(\theta_1|\mathbf{XA},\mathbf{XB},\mathbf{Z},\mathbf{C}_j^{(t)}) \times \pi_1(\theta_1) \nonumber \\
\Prob(\theta_2|\mathbf{XA},\mathbf{XB},\mathbf{Z},\mathbf{C}_j^{(t)}) \propto f_{B_2}(\theta_2|\mathbf{XA},\mathbf{XB},\mathbf{Z},\mathbf{C}_j^{(t)})\times\pi_2(\theta_2)  
\end{eqnarray}

\subsection*{Step 2: Metropolis-Hastings step}
Sampling directly from the distribution of all possible linking permutations is computationally intensive and practically intractable, even within individual blocks, because it requires calculating the likelihood function over all possible permutations. To reduce the computational complexity of this process, we rely on a Metropolis-Hastings algorithm originally proposed by \citet{yingnian-95}. In one iteration of the Metropolis-Hastings step in block $j$, two indices $i_1, i_2 \in \{1,\ldots,I_{A_j}\}$ are chosen at random. From permutation $\mathbf{C}_{j}^{(t)}$ we obtain $\mathbf{C}_{j}^*$ by interchanging the entries at indices $i_1$ and $i_2$ to obtain a new permutation. The ratio between the likelihood of $\mathbf{C}_{j}^*$ and likelihood of $\mathbf{C}_{j}^{(t)}$ is the Metropolis-Hastings acceptance ratio. Formally,
%\Edwin{RG - I removed the following line because we always make file A at least as large as file B because of step 3 so this is not needed}
%While the number of matches in block $j$ is $\min(I_{A_j}, I_{B_j})$, the indices $i_1$ and $i_2$ are in the range $\{1, \ldots, \max(I_{A_j}, I_{B_j})\}$, corresponding to the size of the block $j$, including rows that may be empty in one file. 
\begin{equation}
\label{eq:MH_ratio}
 \mbox{Accept $\mathbf{C}_{j}^*$ with probability } \min \left(1, \frac{L\left(\theta^{(t)}, \mathbf{C}_{j}^* | \mathbf{XA}, \mathbf{XB}, \mathbf{Z}\right)}{L\left(\theta^{(t)}, \mathbf{C}_{j}^{(t)} | \mathbf{XA}, \mathbf{XB}, \mathbf{Z}\right)} \right) \enspace .
\end{equation}

The number of iterations of the Metropolis-Hastings algorithm is set by the user and should be informed by the average block size.

Calculation of the likelihoods ratio in \eqref{eq:MH_ratio} can be simplified to only the terms that include $i_{1}$ and $i_{2}$, because the rest of the likelihood does not change under the swap. When $I_{A_j} > I_{B_j}$, some records in file $A$ will not be linked to records in file $B$, in which case terms corresponding to non-linked records cancel out in the computation of the acceptance ratio. Formally, for the Normal and logistic models described by Equation \eqref{eq:modelExam}, for $i_1$ and $i_2$ in block $j$, assuming they have a link, the likelihood ratio in Equation \eqref{eq:MH_ratio} is 
\begin{flalign*}
\frac{L\left(\theta^{(t)}, \mathbf{C}_{j}^* | \mathbf{XA}, \mathbf{XB}, \mathbf{Z}\right)}{L\left(\theta^{(t)}, \mathbf{C}_{j}^{(t)} | \mathbf{XA}, \mathbf{XB}, \mathbf{Z}\right)} = &&
\end{flalign*}
\begin{eqnarray}
= &\frac{\prod\limits_{i = 1}^{\min\left(I_{A_j}, I_{B_j}\right)} \Bigg[\phi\left(\mathbf{XB}_{C_j^{*}[i],1};\mu_{i}^{(t)}, \sigma^{(t)}\right)\Bigg]\times\Bigg[g^{(t)}_i\left(C_j^{*}\right)^{\mathbf{XB}_{C_j^{*}[i],2}}\left(1-g^{(t)}_i\left(C_j^{*}\right)\right)^{1-\mathbf{XB}_{C_j^{*}[i],2}}\Bigg]}{\prod\limits_{i = 1}^{\min\left(I_{A_j}, I_{B_j}\right)} \Bigg[\phi\left(\mathbf{XB}_{C_j^{(t)}[i],1};\mu_{i}^{(t)}, \sigma^{(t)}\right)\Bigg]\times\Bigg[g^{(t)}_i\left(C_j^{(t)}\right)^{\mathbf{XB}_{C_j^{(t)}[i],2}}\left(1-g^{(t)}_i\left(C_j^{(t)}\right)\right)^{1-\mathbf{XB}_{C_j^{(t)}[i],2}}\Bigg]} \nonumber\\
=& \frac{\prod_{i \in \{i_1,i_2\}} \Bigg[\phi\left(\mathbf{XB}_{C_j^{*}[i]1};\mu_{i}^{(t)}, \sigma^{(t)}\right)\Bigg]\times\Bigg[g^{(t)}_i\left(C_j^{*}\right)^{\mathbf{XB}_{C_j^{*}[i],2}}\left(1-g^{(t)}_i\left(C_j^{*}\right)\right)^{1-\mathbf{XB}_{C_j^{*}[i],2}}\Bigg]}{\prod_{i \in \{i_1,i_2\}} \Bigg[\phi\left(\mathbf{XB}_{C_j^{(t)}[i],1};\mu_{i}^{(t)}, \sigma^{(t)}\right)\Bigg]\times\Bigg[g^{(t)}_i\left(C_j^{(t)}\right)^{\mathbf{XB}_{C_j^{(t)}[i],2}}\left(1-g^{(t)}_i\left(C_j^{(t)}\right)\right)^{1-\mathbf{XB}_{C_j^{(t)}[i],2}}\Bigg]}
\end{eqnarray}
where $\phi(\cdot;\mu^{(t)},\sigma^{(t)})$ is the density of a Normal distribution with mean $\mu$ and standard deviation $\sigma^{(t)}$, $\mu_i^{(t)} = \mathbf{XA}_{i}^{'}\theta^{(t)}_1$ and $g^{(t)}_i(C_{j}) = \text{logit}^{-1}((XB_{C_{j}[i]1},\mathbf{XA}_{i})^{t}\theta^{(t)}_2)$.

\subsection*{Step 3: Imputing missing records in file $A$}
%\Edwin{I'm not sure that this has to be its own "Step." If it should be its own step the title of the section doesn't really make sense anymore. - RG I think that if you read the paper now it might make more sense?}

When file $A$ contains fewer records than file $B$ in  block $j$, we add $I_{B_j}-I_{A_j}$ records of $\mathbf{XA}_{i}$ by sampling with replacement from the $I_{A_j}$ observed $\mathbf{XA}_{i}$ values. This imputation step ensures that for all of the blocks $I_{A_j} \geq I_{B_j}$. 

\begin{comment}
The model implemented in \pkg{gfs\_sampler} considers likelihoods from noise distributed according to Poisson, Normal, or Binomial distributions.  have been factored to increase eliminate duplicated computations. The Poisson, Normal, and Binomial models consider the likelihood following likelihoods, respectively: 

\begin{center}
$ p \left( y _ { 1} ,\dots ,y _ { m } | x _ { 1} ,\dots ,x _ { m } ; \theta \right) = \prod _ { i = 1} ^ { m } \frac { e ^ { y _ { i } \theta ^ { \prime } x _ { i } } e ^ { - e ^ { \theta ^ { \prime } x _ { i } } } } { y _ { i } ! } $

$ \prod _ { i = 1} ^ { n } p \left( y _ { i } | x _ { i } ; \beta _ { 0} ,\beta _ { 1} ,\sigma ^ { 2} \right) = \prod _ { i = 1} ^ { n } \frac { 1} { \sqrt { 2\pi \sigma ^ { 2} } } e ^ { - \frac { \left( y _ { 0} + \beta _ { 1} x _ { i } \right) ^ { 2} } { 2\sigma ^ { 2} } } $

$ L ( \beta | y ) = \prod _ { i = 1} ^ { N } \frac { n _ { i } ! } { y _ { i } ! \left( n _ { i } - y _ { i } \right) ! } \pi _ { i } ^ { y _ { i } } \left( 1- \pi _ { i } \right) ^ { n _ { i } - y _ { i } } $
\end{center} 
\end{comment}

\section{Empirical example}
\label{sc:example}

\begin{comment}
We present the functionality of the \pkg{GFS} (\proglang{R}) and \pkg{gfs\_sampler} (\proglang{Python}) package with an example based on real data. The \pkg{GFS} package can be found at \url{https://github.com/edwinfarley/GFS} and the \pkg{gfs_sampler} package can be found at \url{https://pypi.org/project/gfs-sampler/}. The \pkg{GFS} package can be installed in an \proglang{R} session with a call to \code{install\_github} from the \pkg{devtools} package with the provided url. The \pkg{gfs\_sampler} can be installed using \code{pip} from the Python Package Index. Complete instructions for installing the package can be found at the package links and in the Appendix (Section \ref{sc:appendix}). 
\end{comment}

We present the functionality of the \pkg{GFS} (\proglang{R}) and \pkg{gfs\_sampler} (\proglang{Python}) packages with an example based on real data.

\subsection{The simulated data sets}
The empirical example is based on a subset of 3,200 randomly selected participants from the 2001 Behavioral Risk Factor Surveillance System (BRFSS) data set \citep{brfss}. The participants were randomly selected only among those that had complete data on all of the variables used for the analysis. 

The original data set is partitioned into two files, in order to mimic the input data that is commonly available in file linkage applications. From the available variables in BRFSS, we identify a set of variables that exists in both files and are recorded accurately and variables that are exclusive to each of the files. The variables that appear in both files were used to create blocks. The blocking variables are the individual's sex, state of residence, geographic stratum, and the time of year the entry was recorded, with the year split into four 3-month periods. These chosen blocking variables represent generic measures that could plausibly be shared by two distinct data sets similar to the BRFSS data set. This blocking scheme results in 1725 blocks, 1073 of which contain a single one-to-one link. The remaining 652 blocks with more than 1 record in each file have an average of 3.26 records and a maximum of 24 records. 

The data used can be found at \url{https://github.com/edwinfarley/GFSdata/tree/main/brfss} and loaded into \proglang{Python} with \pkg{pandas}:

\begin{knitrout}
\definecolor{shadecolor}{rgb}{0.969, 0.969, 0.969}\color{fgcolor}\begin{kframe}
\noindent
\ttfamily
\hlstd{}\hlkwa{import\ }\hlstd{pandas\ }\hlkwa{as\ }\hlstd{pd}\hspace*{\fill}\\
\hlstd{samples\ }\hlopt{=\ }\hlstd{pd}\hlopt{.}\hlstd{}\hlkwd{read\textunderscore csv}\hlstd{}\hlopt{(}\hlstd{}\hlstr{"https://raw.githubusercontent.com/edwinfarley/"}\hlstd{\ }\hspace*{\fill}\\
\hlstd{\ }\hlopt{+\ }\hlstd{}\hlstr{"GFSdata/main/brfss/samples.csv"}\hlstd{}\hlopt{)}\hspace*{\fill}\\
\hlstd{samples1\ }\hlopt{=\ }\hlstd{pd}\hlopt{.}\hlstd{}\hlkwd{read\textunderscore csv}\hlstd{}\hlopt{(}\hlstd{}\hlstr{"https://raw.githubusercontent.com/edwinfarley/"}\hlstd{\ }\hspace*{\fill}\\
\hlstd{\ }\hlopt{+\ }\hlstd{}\hlstr{"GFSdata/main/brfss/samples1.csv"}\hlstd{}\hlopt{)}\hspace*{\fill}\\
\hlstd{samples2\ }\hlopt{=\ }\hlstd{pd}\hlopt{.}\hlstd{}\hlkwd{read\textunderscore csv}\hlstd{}\hlopt{(}\hlstd{}\hlstr{"https://raw.githubusercontent.com/edwinfarley/"}\hlstd{\ }\hspace*{\fill}\\
\hlstd{\ }\hlopt{+\ }\hlstd{}\hlstr{"GFSdata/main/brfss/samples2.csv"}\hlstd{}\hlopt{)}\hlstd{}\hspace*{\fill}
\mbox{}
\normalfont
\end{kframe}
\end{knitrout}

The variables that were selected to be included in this analysis include individuals' general health, physical health, mental health, age, alcohol consumption, weight, and whether they suffer from asthma. A summary of the data using \proglang{Python} is:

\begin{knitrout}
\definecolor{shadecolor}{rgb}{0.969, 0.969, 0.969}\color{fgcolor}\begin{kframe}
\noindent
\ttfamily
\hlstd{samples}\hspace*{\fill}
\mbox{}
\normalfont

\begin{verbatim}
##    GENHLTH  PHYSHLTH  MENTHLTH  AGE  ALCDAY  WEIGHT  ASTHMA  block
## 0        3        10        20   36     215     202       1      1
## 1        3        15         6   57     207     260       1      2
## 2        3         2         5   24     102     175       0      3
## 3        4        20        20   63     204     275       0      4
## 4        3         2        10   37     101     161       1      5
## ...    ...       ...       ...  ...     ...     ...     ...    ...
## 3195     4        30        30   54     230     100       1   1721
## 3196     3        15        15   39     201     200       1   1722
## 3197     4         1         1   29     212     176       1   1723
## 3198     3         1         3   27     201     130       1   1724
## 3199     4         6         3   62     201     130       1   1725

\end{verbatim}
\end{kframe}
\end{knitrout}

The variables in the first file include the individuals' weight, physical health, mental health, and age. A summary of the data in the first file is:

\begin{knitrout}
\definecolor{shadecolor}{rgb}{0.969, 0.969, 0.969}\color{fgcolor}\begin{kframe}
\noindent
\ttfamily
\hlstd{samples1}\hspace*{\fill}
\mbox{}
\normalfont

\begin{verbatim}
##    WEIGHT  PHYSHLTH  MENTHLTH  AGE  block
## 0     202        10        20   36      1
## 1     260        15         6   57      2
## 2     175         2         5   24      3
## 3     275        20        20   63      4
## 4     161         2        10   37      5
## ...   ...       ...       ...  ...    ...
## 3195  100        30        30   54   1721
## 3196  200        15        15   39   1722
## 3197  176         1         1   29   1723
## 3198  130         1         3   27   1724
## 3199  130         6         3   62   1725
\end{verbatim}
\end{kframe}
\end{knitrout}

The variables in the second file include the individuals' general health, alcohol consumption, and an indicator for asthma. A summary of the data in the second file is:

\begin{knitrout}
\definecolor{shadecolor}{rgb}{0.969, 0.969, 0.969}\color{fgcolor}\begin{kframe}
\noindent
\ttfamily
\hlstd{samples2}\hspace*{\fill}
\mbox{}
\normalfont

\begin{verbatim}
##   GENHLTH  ALCDAY  ASTHMA  block
## 0       2     101       0    624
## 1       4     228       0    514
## 2       3     201       1   1120
## 3       4     201       1    272
## 4       2     103       0    145
## ...   ...     ...     ...    ...
## 3195    2     101       1    776
## 3196    2     201       1   1705
## 3197    3     204       0   1328
## 3198    2     201       1   1059
## 3199    1     105       1   1288
\end{verbatim}
\end{kframe}
\end{knitrout}

The block column present in both files indicates the blocks to which each record belongs, with similar indexing in both data sets.

\subsection{Linkage models}
\label{subs:samplingMod}

We applied the record linkage algorithm under three separate $f_{B|A}$ models. The first model was a linear regression model with the general health measure as the response variable (\code{GENHLTH}), and physical health (\code{PHYSHLTH}) and mental health (\code{MENTHLTH}) as explanatory variables. The second model was a logistic regression model with the asthma indicator, \code{ASTHMA}, as the response variable, and \code{PHYSHLTH}, \code{WEIGHT}, and \code{AGE} as the explanatory variables. The third model relies on both models jointly to link records from both files.

\begin{comment}
In practice, when working with two distinct data sets it is not possible to perform any sort of correlation analysis between data sets to identify good candidates for the response variable and its covariates. Instead, it is often best to use as many covariates as possible from one data set and select a response variable with decent variance. We elected for a simple formula for this example, with columns that could reasonably be assumed to be correlated. Note that the sampling procedure itself is independent from the statistic for which the linkage is being performed.
\end{comment}

\subsection{Single generalized linear models}
\label{subs:SingRegMod}

To sample from the joint posterior distribution of the linkage structure and the parameters in \proglang{Python} we use the \code{sample()} function from the \pkg{gfs\_sampler} package. Its first two input parameters are the data sets to be linked. The third input (\code{formula\_array}) is an array of string formulas. The format of each formula is similar to the expression of a formula in \proglang{R}. Specifically, the form is $y \sim model$, where $y$ is the response variable and linear predictors are specified symbolically by $model$. The format of $model$ consists of a series of column names in either file $A$ or file $B$ separated by $+$ operators. The fourth parameter (\code{family\_array}) defines the type of regression associated with each formula, where possible values include \code{"Normal"} for a linear regression model, \code{"Logistic"} for a logistic regression model and \code{"Poisson"} for a log-linear model. The fifth parameter (\code{M}) defines the number of linkage permutations that will be produced by the sampling algorithm. The sixth parameter (\code{I}) defines the number of iterations used to sample the parameters, $\theta$, within each iteration of the sampling algorithm described in Section \ref{sc:sampling}. The seventh parameter (\code{t}) defines the number of iterations of the Metropolis-Hastings step will be used to sample new permutations within each iteration of the sampling algorithm in Section \ref{sc:sampling}. The total number of iterations is the product of this value and the number of records in each block. The eighth parameter (\code{burnin}) is the number of burn-in iterations of the sampling algorithm. The ninth parameter (\code{interval}) is the thinning interval, which is the number of iterations discarded between each saved sample of the sampling algorithm. 

\begin{knitrout}
\definecolor{shadecolor}{rgb}{0.969, 0.969, 0.969}\color{fgcolor}\begin{kframe}
\noindent
\ttfamily
\hlstd{}\hlslc{\#\ M:\ number\ of\ samples}\hspace*{\fill}\\
\hlstd{M\ }\hlopt{=\ }\hlstd{}\hlnum{10}\hspace*{\fill}\\
\hlstd{}\hlslc{\#\ I:\ number\ of\ iterations\ for\ sampling\ regression\ coefficients}\hspace*{\fill}\\
\hlstd{I\ }\hlopt{=\ }\hlstd{}\hlnum{50}\hspace*{\fill}\\
\hlstd{}\hlslc{\#\ t:\ Metropolis{-}Hastings\ samples\ multiplier}\hspace*{\fill}\\
\hlstd{t\ }\hlopt{=\ }\hlstd{}\hlnum{5}\hspace*{\fill}\\
\hlstd{}\hlslc{\#\ burnin:\ number\ of\ burn{-}in\ samples}\hspace*{\fill}\\
\hlstd{burnin\ }\hlopt{=\ }\hlstd{}\hlnum{200}\hspace*{\fill}\\
\hlstd{}\hlslc{\#\ interval:\ number\ of\ iterations\ between\ samples,\ after\ burn{-}in}\hspace*{\fill}\\
\hlstd{interval\ }\hlopt{=\ }\hlstd{}\hlnum{20}\hlstd{}\hspace*{\fill}
\mbox{}
\normalfont
\end{kframe}
\end{knitrout}

The following \proglang{Python} command is used to link the two data sets using a linear regression model (\code{"Normal"}):

\begin{knitrout}
\definecolor{shadecolor}{rgb}{0.969, 0.969, 0.969}\color{fgcolor}\begin{kframe}
\noindent
\ttfamily
\hlstd{}\hlkwa{import\ }\hlstd{gfs\textunderscore sampler\ }\hlkwa{as\ }\hlstd{gfs}\hspace*{\fill}\\
\hlstd{}\hspace*{\fill}\\
\hlstd{P\textunderscore samples\ }\hlopt{=\ }\hlstd{gfs}\hlopt{.}\hlstd{}\hlkwd{sample}\hlstd{}\hlopt{(}\hlstd{samples1}\hlopt{,\ }\hlstd{samples2}\hlopt{,}\hspace*{\fill}\\
\hlstd{\ }\hlopt{{[}}\hlstd{}\hlstr{"GENHLTH$\sim$PHYSHLTH+MENTHLTH"}\hlstd{}\hlopt{{]},\ {[}}\hlstd{}\hlstr{"Normal"}\hlstd{}\hlopt{{]},\ }\hspace*{\fill}\\
\hlstd{\ M}\hlopt{,\ }\hlstd{I}\hlopt{,\ }\hlstd{t}\hlopt{,\ }\hlstd{burnin}\hlopt{,\ }\hlstd{interval}\hspace*{\fill}\\
\hlstd{}\hlopt{)}\hlstd{}\hspace*{\fill}
\mbox{}
\normalfont
\end{kframe}
\end{knitrout}

\begin{comment}
We want to obtain 10 permutations to link the data sets and estimate statistics over the hypothetical complete data set, which would be unknown in a real application. Based on the size of the data set, we set the number of iterations in sampling the regression parameters to 25, and we set the number of iterations in the Metropolis-Hastings step to be 5 times the number of rows in each block. The number of burn-in iterations should be set based on the size of the data set and complexity of the model. We aimed to consider a simple example, so there was little to gain from increasing the number of samples computed beyond a level that would ensure we started to show convergence in the chain. We set the number of burn-in samples to 750 and the sampling interval to 50 iterations so that individual samples could reasonably be assumed to be independent. Parameter choice is not an exact science, but these values ensure that we will adequately explore different permutations in the larger blocks without overdoing it on smaller blocks.
\end{comment}

The following \proglang{Python} command is used to link the two data sets using a logistic regression model (\code{"Logistic"}):

\begin{knitrout}
\definecolor{shadecolor}{rgb}{0.969, 0.969, 0.969}\color{fgcolor}\begin{kframe}
\noindent
\ttfamily
\hlstd{P\textunderscore log\textunderscore samples\ }\hlopt{=\ }\hlstd{gfs}\hlopt{.}\hlstd{}\hlkwd{sample}\hlstd{}\hlopt{(}\hlstd{samples1}\hlopt{,\ }\hlstd{samples2}\hlopt{,}\hspace*{\fill}\\
\hlstd{\ }\hlopt{{[}}\hlstd{}\hlstr{"ASTHMA$\sim$PHYSHLTH+AGE+WEIGHT"}\hlstd{}\hlopt{{]},\ {[}}\hlstd{}\hlstr{"Logistic"}\hlstd{}\hlopt{{]},\ }\hspace*{\fill}\\
\hlstd{\ M}\hlopt{,\ }\hlstd{I}\hlopt{,\ }\hlstd{t}\hlopt{,\ }\hlstd{burnin}\hlopt{,\ }\hlstd{interval}\hspace*{\fill}\\
\hlstd{}\hlopt{)}\hlstd{}\hspace*{\fill}
\mbox{}
\normalfont
\end{kframe}
\end{knitrout}

The return value of these calls is a data frame of size 3200 $\times$ \code{M}, where each column contains row indices defining a sampled permutation $\mathbf{C}^{(m)}, m \in \{1,\ldots,M\}$. The permutation represents which rows from the second data set  are linked to records of the first data set. This is an efficient manner to store the permutations without saving multiple copies of the two files. We provide a function to generate complete data sets from the two files and a given permutation. Point estimates can be obtained by computing the estimates of interest within each linked data set and averaging these values. Interval estimates for these quantities can be obtained using common multiple imputation combination rules \citep{rubin2004multiple}.

\subsection{Multiple generalized linear models}
\label{subs:JointModel}

For bivariate Normal records that are partitioned across two files, it was shown that the expected gain in information using the proposed algorithm increases as the correlation between the variables increases \citep{Gutman}. One way to increase the correlation between records is to add more response and explanatory variables. Adding models with different response variables can improve the point and interval estimates of the parameters of interest, as well as increase the number of correctly identified links. 

The \pkg{gfs\_sampler} package supports the use of multiple linkage models to inform the linkage procedure. We consider combining the linear regression model and the logistic regression model described in Section \ref{subs:SingRegMod}. The \code{Python} command for running the two models jointly is:

\begin{knitrout}
\definecolor{shadecolor}{rgb}{0.969, 0.969, 0.969}\color{fgcolor}\begin{kframe}
\noindent
\ttfamily
\hlstd{P\textunderscore joint\textunderscore samples\ }\hlopt{=\ }\hlstd{gfs}\hlopt{.}\hlstd{}\hlkwd{sample}\hlstd{}\hlopt{(}\hlstd{samples1}\hlopt{,\ }\hlstd{samples2}\hlopt{,}\hspace*{\fill}\\
\hlstd{\ }\hlopt{{[}}\hlstd{}\hlstr{"GENHLTH$\sim$PHYSHLTH+MENTHLTH"}\hlstd{}\hlopt{,\ }\hlstd{}\hlstr{"ASTHMA$\sim$PHYSHLTH+AGE+WEIGHT+GENHLTH"}\hlstd{}\hlopt{{]},\ }\hspace*{\fill}\\
\hlstd{\ }\hlopt{{[}}\hlstd{}\hlstr{"Normal"}\hlstd{}\hlopt{,\ }\hlstd{}\hlstr{"Logistic"}\hlstd{}\hlopt{{]},\ }\hspace*{\fill}\\
\hlstd{\ M}\hlopt{,\ }\hlstd{I}\hlopt{,\ }\hlstd{t}\hlopt{,\ }\hlstd{burnin}\hlopt{,\ }\hlstd{interval}\hspace*{\fill}\\
\hlstd{}\hlopt{)}\hlstd{}\hspace*{\fill}
\mbox{}
\normalfont
\end{kframe}
\end{knitrout}

\subsection{Using linked data sets}

To demonstrate possible analyses using the linked data sets, we describe regression analyses using variables that are exclusive to one of the two data sets. Our analysis consists of generalized linear models, in which the response variable is from one file and the explanatory variable is from the other file. To obtain a linked data set based on a sampled permutation, we create a copy of the second data set, where rows are reordered according to the indices in the permutation. This is implemented by indexing the records of \code{samples2} with an array corresponding to permutation \code{m}, which is one of the columns obtained as output of the command \code{permute\_inputs}. This results in records from file \code{samples2} that can be horizontally concatenated to records in file \code{samples1}. In \proglang{Python} this is implemented with the following command: 

\begin{knitrout}
\definecolor{shadecolor}{rgb}{0.969, 0.969, 0.969}\color{fgcolor}\begin{kframe}
\noindent
\ttfamily
\hlstd{samples\textunderscore linked\ }\hlopt{=\ }\hlstd{gfs}\hlopt{.}\hlstd{}\hlkwd{apply\textunderscore permutation}\hlstd{}\hlopt{(}\hlstd{samples1}\hlopt{,\ }\hlstd{samples2}\hlopt{,\ }\hlstd{m}\hlopt{)}\hlstd{}\hspace*{\fill}
\mbox{}
\normalfont
\end{kframe}
\end{knitrout}

Applying this function to each of the sampled permutations will result in multiple fully linked data sets. The statistic of interest can be calculated using each of these data sets separately. The individual results are then combined using the common multiple imputation combination rules to obtain point and interval estimates. 

Because \proglang{R} provides larger set of possible statistical analyses procedures, we save the permutation results to comma delimited files and load them in \proglang{R} for further analysis.

\begin{knitrout}
\definecolor{shadecolor}{rgb}{0.969, 0.969, 0.969}\color{fgcolor}\begin{kframe}
\noindent
\ttfamily
\hlstd{P\textunderscore samples}\hlopt{.}\hlstd{}\hlkwd{to\textunderscore csv}\hlstd{}\hlopt{(}\hlstd{}\hlstr{"P.csv"}\hlstd{}\hlopt{)}\hspace*{\fill}\\
\hlstd{P\textunderscore log\textunderscore samples}\hlopt{.}\hlstd{}\hlkwd{to\textunderscore csv}\hlstd{}\hlopt{(}\hlstd{}\hlstr{"P\textunderscore log.csv"}\hlstd{}\hlopt{)}\hspace*{\fill}\\
\hlstd{P\textunderscore joint\textunderscore samples}\hlopt{.}\hlstd{}\hlkwd{to\textunderscore csv}\hlstd{}\hlopt{(}\hlstd{}\hlstr{"P\textunderscore joint.csv"}\hlstd{}\hlopt{)}\hlstd{}\hspace*{\fill}
\mbox{}
\normalfont
\end{kframe}
\end{knitrout}

\subsection{Results}

We present the estimates for the slopes in generalized linear models with one explanatory variable using two linkage procedures. The first procedure samples permutations from a uniform distribution over all possible permutations. The second procedure is the proposed linking procedure. For the proposed procedure we examined three different models described in Sections \ref{subs:SingRegMod} and \ref{subs:JointModel}. We compare the slope estimates from the two different linkage procedures to the slope estimates from a data set based on the true linkage.

To analyze the linked data set, we first load the two separate files into the \proglang{R} environment.

\begin{knitrout}
\definecolor{shadecolor}{rgb}{0.969, 0.969, 0.969}\color{fgcolor}\begin{kframe}
\begin{alltt}
\hlstd{> }\hlstd{baseURL} \hlkwb{=} \hlstr{"https://raw.githubusercontent.com/edwinfarley/GFSdata/"}
\hlstd{> }\hlstd{samples1} \hlkwb{=} \hlkwd{read.csv}\hlstd{(}\hlkwd{paste}\hlstd{(baseURL,} \hlstr{"main/brfss/samples1.csv"}\hlstd{,} \hlkwc{sep} \hlstd{=} \hlstr{""}\hlstd{))}
\hlstd{> }\hlstd{samples2} \hlkwb{=} \hlkwd{read.csv}\hlstd{(}\hlkwd{paste}\hlstd{(baseURL,} \hlstr{"main/brfss/samples2.csv"}\hlstd{,} \hlkwc{sep} \hlstd{=} \hlstr{""}\hlstd{))}
\end{alltt}
\end{kframe}
\end{knitrout}

Because \proglang{Python} uses 0-based indexing, we add 1 to the results obtained from the \pkg{gfs\_sampler} after loading the permutations to \proglang{R}.

\begin{knitrout}
\definecolor{shadecolor}{rgb}{0.969, 0.969, 0.969}\color{fgcolor}\begin{kframe}
\begin{alltt}
\hlstd{> }\hlstd{P} \hlkwb{=} \hlkwd{read.csv}\hlstd{(}\hlstr{"P.csv"}\hlstd{)} \hlopt{+} \hlnum{1}
\hlstd{> }\hlstd{P_log} \hlkwb{=} \hlkwd{read.csv}\hlstd{(}\hlstr{"P_log.csv"}\hlstd{)} \hlopt{+} \hlnum{1}
\hlstd{> }\hlstd{P_joint} \hlkwb{=} \hlkwd{read.csv}\hlstd{(}\hlstr{"P_joint.csv"}\hlstd{)} \hlopt{+} \hlnum{1}
\hlstd{ }
\hlstd{> }\hlstd{P}\hlopt{$}\hlstd{X} \hlkwb{=} \hlkwa{NULL}
\hlstd{> }\hlstd{P_log}\hlopt{$}\hlstd{X} \hlkwb{=} \hlkwa{NULL}
\hlstd{> }\hlstd{P_joint}\hlopt{$}\hlstd{X} \hlkwb{=} \hlkwa{NULL}
\end{alltt}
\end{kframe}
\end{knitrout}

We provide the code for estimating the slope of linear regression model with \code{GENHLTH} as the dependent variable and \code{PHYSHLTH} as the independent variable. This code can be easily adjusted to obtain estimates for other generalized linear models. 

\begin{knitrout}
\definecolor{shadecolor}{rgb}{0.969, 0.969, 0.969}\color{fgcolor}\begin{kframe}
\begin{alltt}
\hlcom{# Confidence level for interval}
\hlstd{a} \hlkwb{=} \hlnum{0.95}
\hlstd{coeffs} \hlkwb{=} \hlkwd{numeric}\hlstd{(}\hlkwd{ncol}\hlstd{(P))}
\hlstd{stderrs} \hlkwb{=} \hlkwd{numeric}\hlstd{(}\hlkwd{ncol}\hlstd{(P))}
\hlstd{Z} \hlkwb{=} \hlkwd{qnorm}\hlstd{(a} \hlopt{+} \hlstd{((}\hlnum{1} \hlopt{-} \hlstd{a)} \hlopt{/} \hlnum{2}\hlstd{))}
\hlcom{# Set model for to compute coefficient}
\hlstd{model} \hlkwb{=} \hlstr{"GENHLTH~PHYSHLTH"}

\hlcom{# Apply each permutation}
\hlkwa{for}\hlstd{(i} \hlkwa{in} \hlnum{1}\hlopt{:}\hlkwd{ncol}\hlstd{(P))\{}
        \hlstd{p} \hlkwb{=} \hlkwd{as.vector}\hlstd{(P[, i])}
        \hlcom{# There is a build_permutation function in the GFS R package,}
        \hlcom{#  but in this case, because blocks are consistent sizes in both files,}
        \hlcom{# we can simply index by the permutation. }
        \hlstd{samples2_unshuffled} \hlkwb{=} \hlstd{samples2[p, ]}
        \hlstd{samples_i} \hlkwb{=} \hlkwd{cbind}\hlstd{(samples1, samples2_unshuffled)}
        \hlstd{samples_i} \hlkwb{=} \hlstd{samples_i[}\hlkwd{complete.cases}\hlstd{(samples_i), ]}
        \hlstd{linreg} \hlkwb{=} \hlkwd{summary}\hlstd{(}\hlkwd{lm}\hlstd{(model,} \hlkwc{data} \hlstd{= samples_i))}\hlopt{$}\hlstd{coefficients}
        \hlstd{coeffs[i]} \hlkwb{=} \hlstd{linreg[}\hlnum{2}\hlstd{,} \hlnum{1}\hlstd{]}
        \hlstd{stderrs[i]} \hlkwb{=} \hlstd{linreg[}\hlnum{2}\hlstd{,} \hlnum{2}\hlstd{]}
\hlstd{\}}

\hlcom{#Use Multple Imputation rules to obtain estimate and total variance}
\hlstd{estimate} \hlkwb{=} \hlkwd{mean}\hlstd{(coeffs)}
\hlstd{total_var} \hlkwb{=} \hlstd{((}\hlnum{1} \hlopt{+} \hlstd{(}\hlnum{1} \hlopt{/} \hlkwd{ncol}\hlstd{(P)))} \hlopt{*} \hlkwd{var}\hlstd{(coeffs))} \hlopt{+} \hlkwd{mean}\hlstd{(stderrs}\hlopt{^}\hlnum{2}\hlstd{)}

\hlcom{# Confidence interval for estimate}
\hlstd{interval} \hlkwb{=} \hlkwd{c}\hlstd{(estimate} \hlopt{-} \hlstd{Z} \hlopt{*} \hlkwd{sqrt}\hlstd{(total_var), estimate} \hlopt{+} \hlstd{Z} \hlopt{*} \hlkwd{sqrt}\hlstd{(total_var))}
\end{alltt}
\end{kframe}
\end{knitrout}

The Normal approximation used to calculate the interval estimates is based on large samples and when the number of imputation approaches infinity \citep{rubin2004multiple}. Our sample is relatively large, but the number of imputation is finite. In large samples with finite number of imputations a reference t-distribution was proposed to calculate interval estimates. \citet{SmallSampleMI} derived a reference t-distribution with degrees of freedom estimates that perform well moderate samples. The code below demonstrates interval estimates using the t-distributions with the degress of freedom estimates proposed by \citet{SmallSampleMI}. 

\begin{knitrout}
\definecolor{shadecolor}{rgb}{0.969, 0.969, 0.969}\color{fgcolor}\begin{kframe}
\begin{alltt}
\hlstd{rd} \hlkwb{=} \hlstd{((}\hlnum{1} \hlopt{+} \hlstd{(}\hlnum{1} \hlopt{/} \hlkwd{ncol}\hlstd{(P)))} \hlopt{*} \hlkwd{var}\hlstd{(coeffs))} \hlopt{/} \hlkwd{mean}\hlstd{(stderrs}\hlopt{^}\hlnum{2}\hlstd{)}
\hlstd{v} \hlkwb{=} \hlstd{(}\hlkwd{ncol}\hlstd{(P)} \hlopt{-} \hlnum{1}\hlstd{)} \hlopt{*} \hlstd{(}\hlnum{1} \hlopt{+} \hlstd{rd}\hlopt{^-}\hlnum{1}\hlstd{)}\hlopt{^}\hlnum{2}
\hlstd{vcom} \hlkwb{=} \hlkwd{dim}\hlstd{(samples1)[}\hlnum{1}\hlstd{]} \hlopt{-} \hlnum{2}
\hlstd{vobs} \hlkwb{=} \hlstd{(}\hlnum{1} \hlopt{-} \hlstd{rd)} \hlopt{*} \hlstd{((vcom} \hlopt{+} \hlnum{1}\hlstd{)} \hlopt{/} \hlstd{(vcom} \hlopt{+} \hlnum{3}\hlstd{))} \hlopt{*} \hlstd{vcom}
\hlstd{vstar} \hlkwb{=} \hlstd{(v}\hlopt{^-}\hlnum{1} \hlopt{+} \hlstd{vobs}\hlopt{^-}\hlnum{1}\hlstd{)}\hlopt{^-}\hlnum{1}

\hlstd{t} \hlkwb{=} \hlkwd{qt}\hlstd{(a} \hlopt{+} \hlstd{((}\hlnum{1} \hlopt{-} \hlstd{a)} \hlopt{/} \hlnum{2}\hlstd{), vstar)}
\hlcom{# Confidence interval for estimate according to t-score}
\hlstd{interval} \hlkwb{=} \hlkwd{c}\hlstd{(estimate} \hlopt{-} \hlstd{t} \hlopt{*} \hlkwd{sqrt}\hlstd{(total_var), estimate} \hlopt{+} \hlstd{t} \hlopt{*} \hlkwd{sqrt}\hlstd{(total_var))}
\end{alltt}
\end{kframe}
\end{knitrout}

\subsubsection{Linear regression linkage model}
\label{subs:RegLin}

\begin{figure}[htbp]
\centering
\includegraphics[width=0.45\textwidth]{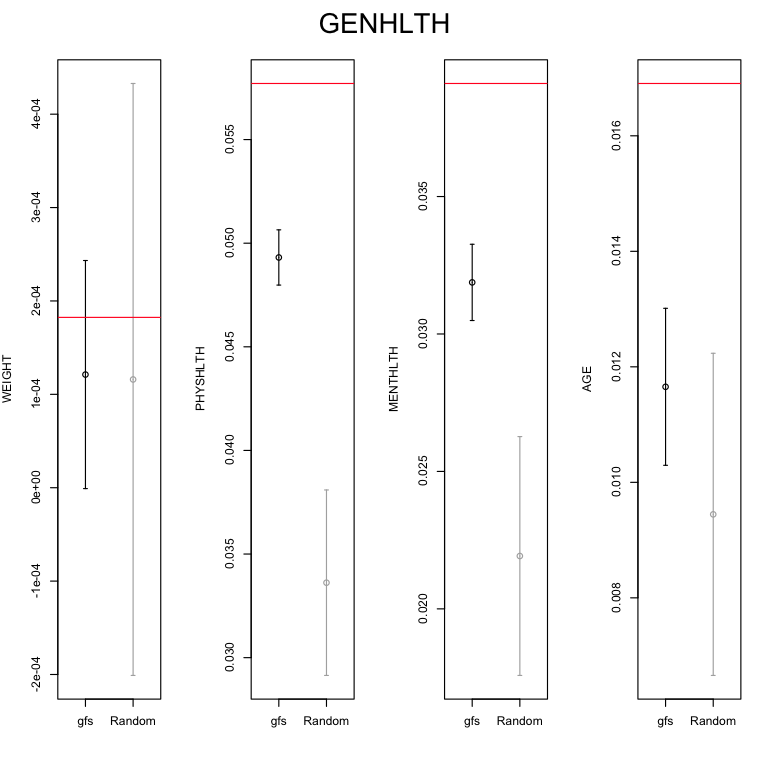} \includegraphics[width=0.45\textwidth]{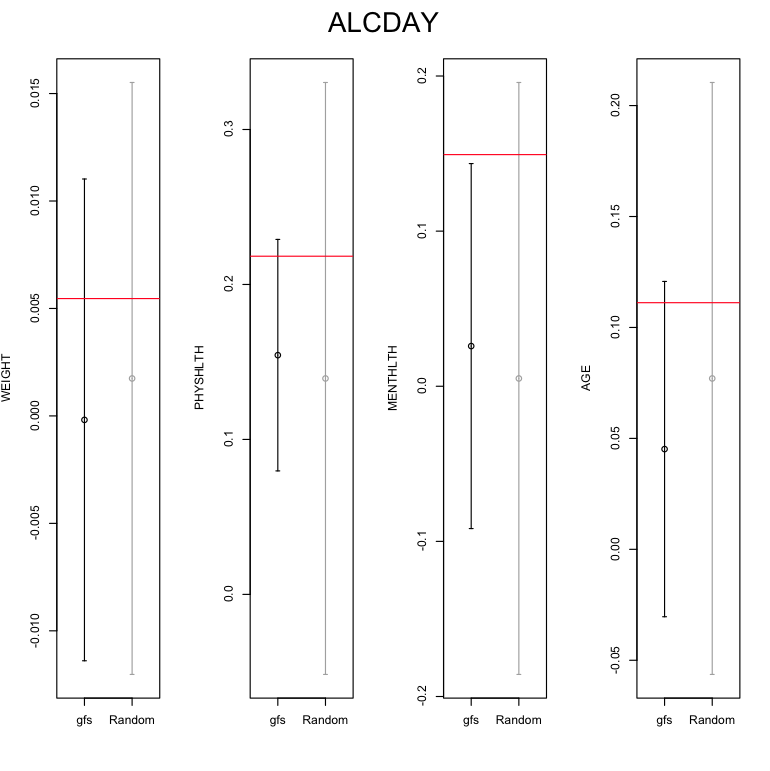}\\
\includegraphics[width=0.45\textwidth]{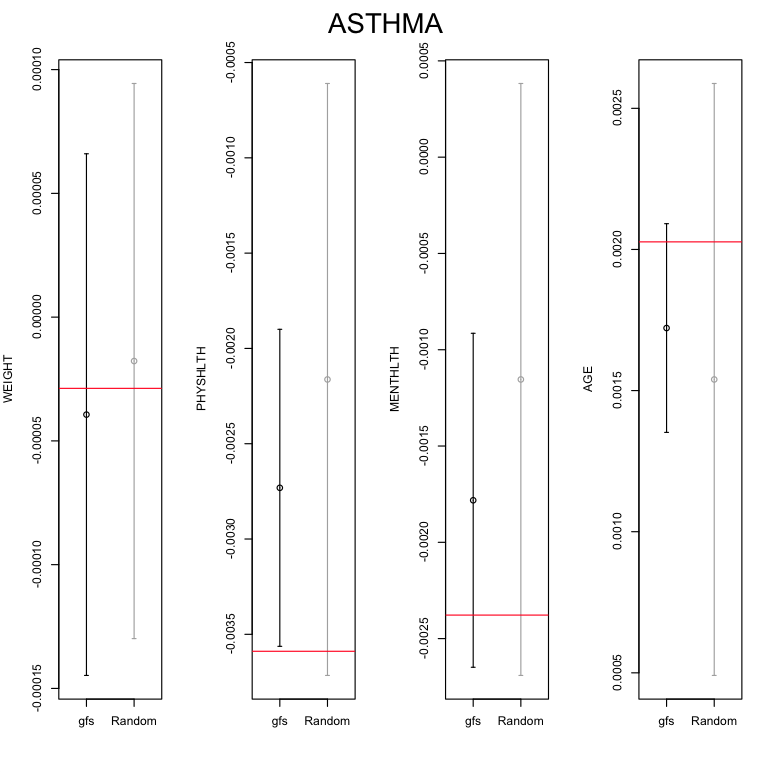}
\caption{\code{GENHLTH}, \code{ALCDAY}, and \code{ASTHMA} under the Normal model.}
\label{fig:normal_plots}
\end{figure}

Figure \ref{fig:normal_plots} shows the results of multiple generalized linear models with one covariate when linkage is implemented with a linear regression model. The title of each set of plots is the response variable, and the plots within each set show the point and interval estimates of the slopes for the covariate indicated on the y-axis. In each plot, the red line depicts the value of the slope calculated with the complete data set, the black and grey intervals show the 95\% confidence intervals under linkage permutations sampled using the proposed method and those sampled under random permutations, respectively.

The proposed method generally reduces bias in point estimates compared to the randomly sampled permutations. This is most apparent when \code{GENHLTH} is the response variable. The proposed method results in significantly smaller sampling variance than is observed with randomly selected permutations. Moreover, the interval estimates of both methods display similar coverage of the true slope. This shows the gain in accuracy and precision using the proposed method.

\subsubsection{Logistic regression linkage model}

\begin{figure}[t]
\centering
\includegraphics[width=0.45\textwidth]{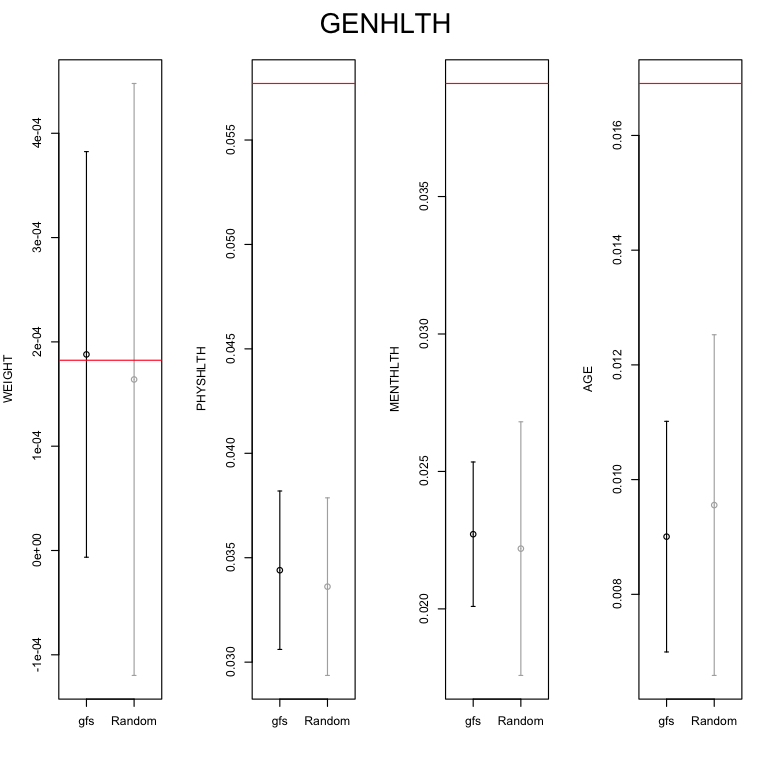} \includegraphics[width=0.45\textwidth]{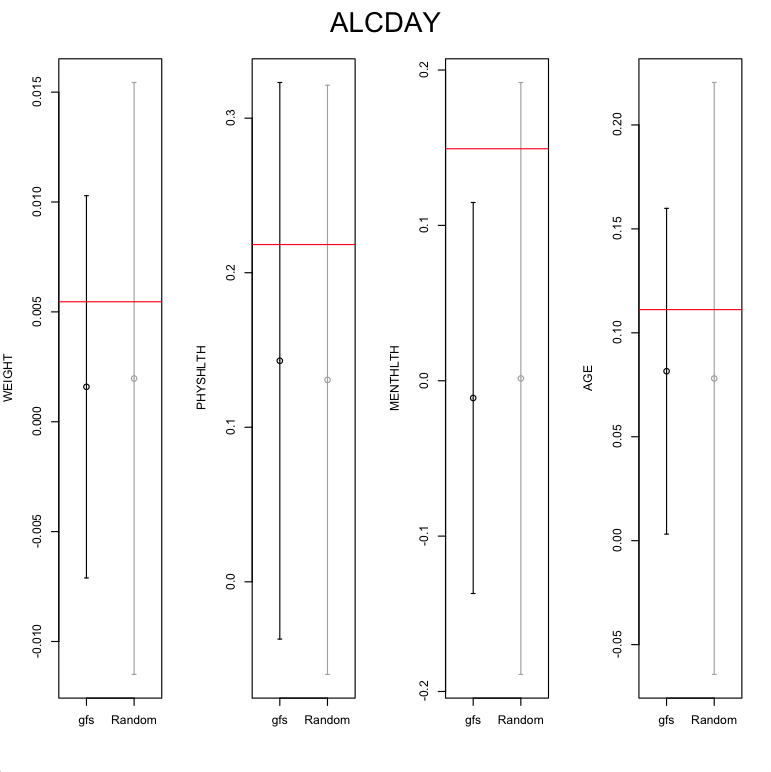}\\
\includegraphics[width=0.45\textwidth]{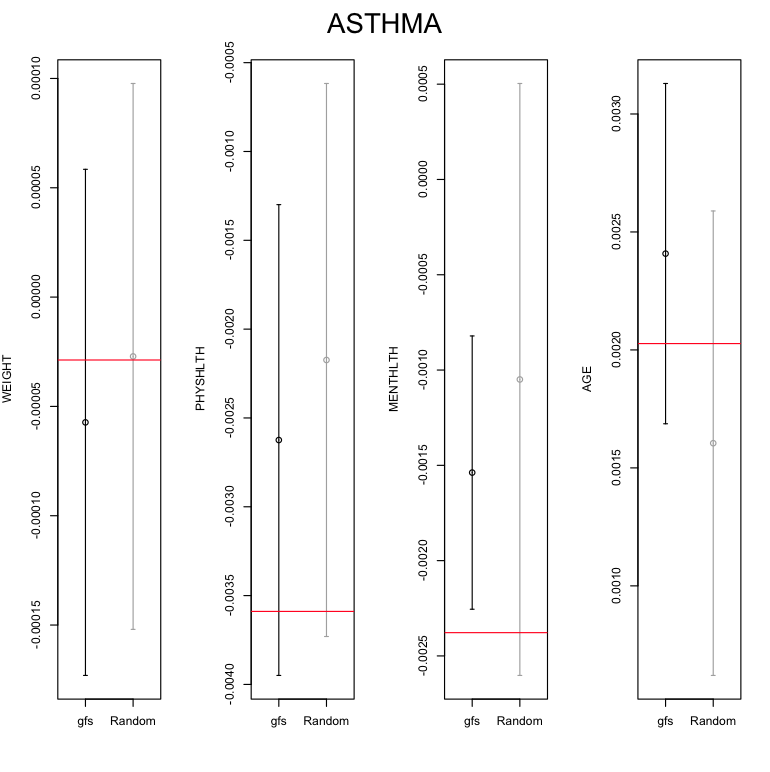}
\caption{\code{GENHLTH}, \code{ALCDAY}, and \code{ASTHMA} under the Logistic model.}
\label{fig:logistic_plots}
\end{figure}

Figure \ref{fig:logistic_plots} depicts the results for different generalized linear models with a single covariate when the linkage method is based on a logistic regression model. The linkage using the proposed method for the \code{GENHLTH} and the \code{ALCDAY} response variables are similar to the random permutations. For the response variable \code{ASTHMA}, we observe slightly lower bias with most explanatory variables and smaller sampling variance. An explanation for the worse performance of the logistic model compared to the regression model in the previous section is that there are 984 records that belong to blocks where the value of \code{ASTHMA} is identical for all records in the block. The coefficients of the logistic regression model are sampled at each MCMC iteration over the entire data set; however, when sampling a new permutation, if all the binary response values in a block are the same, all the block's linkage permutations have identical likelihood. Therefore, in any block with identical binary response values the proposed procedure samples linkage permutations uniformly. 

\begin{figure}[htbp]
\centering
\includegraphics[width=0.45\textwidth]{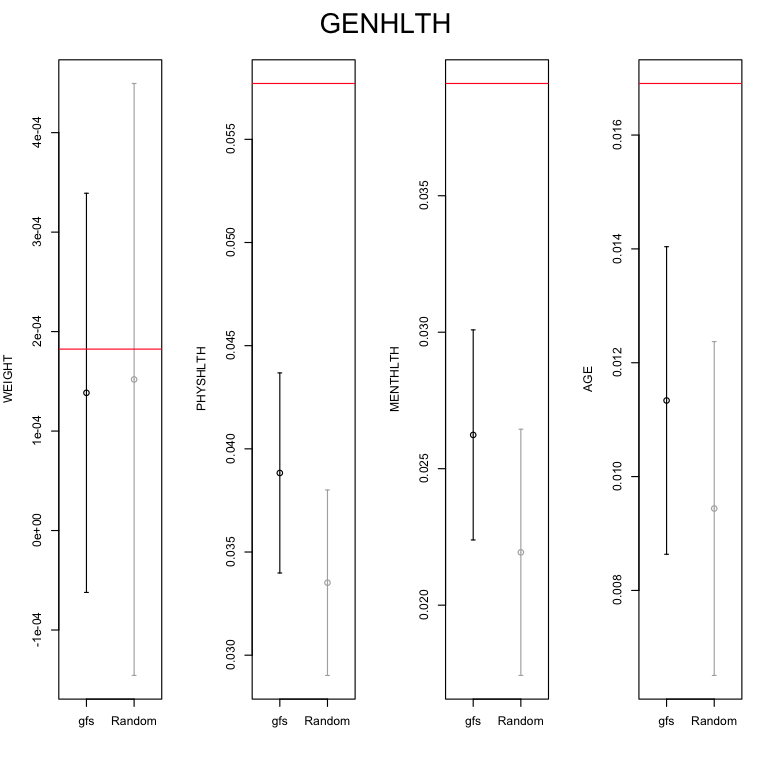} \includegraphics[width=0.45\textwidth]{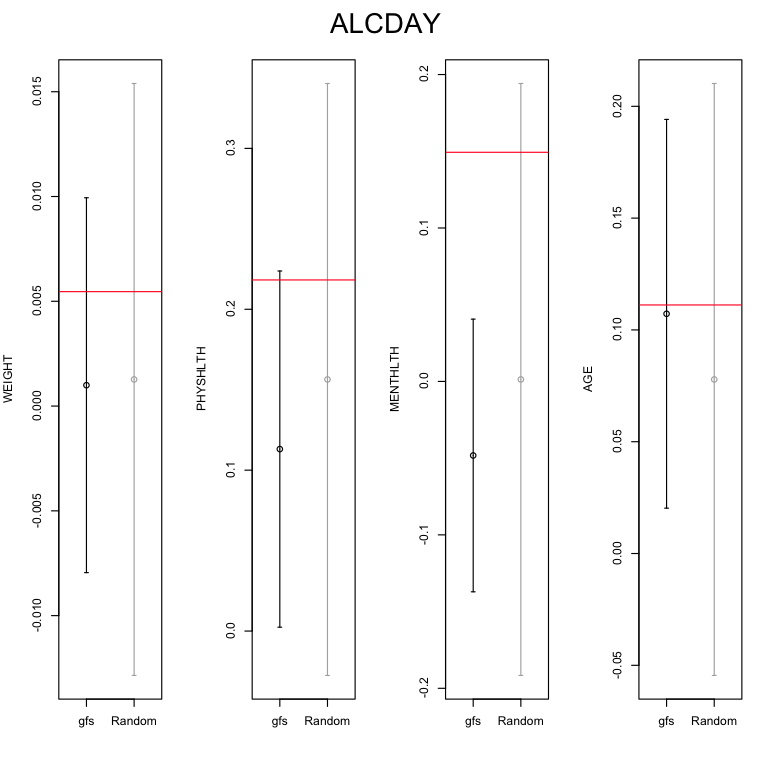}\\
\includegraphics[width=0.45\textwidth]{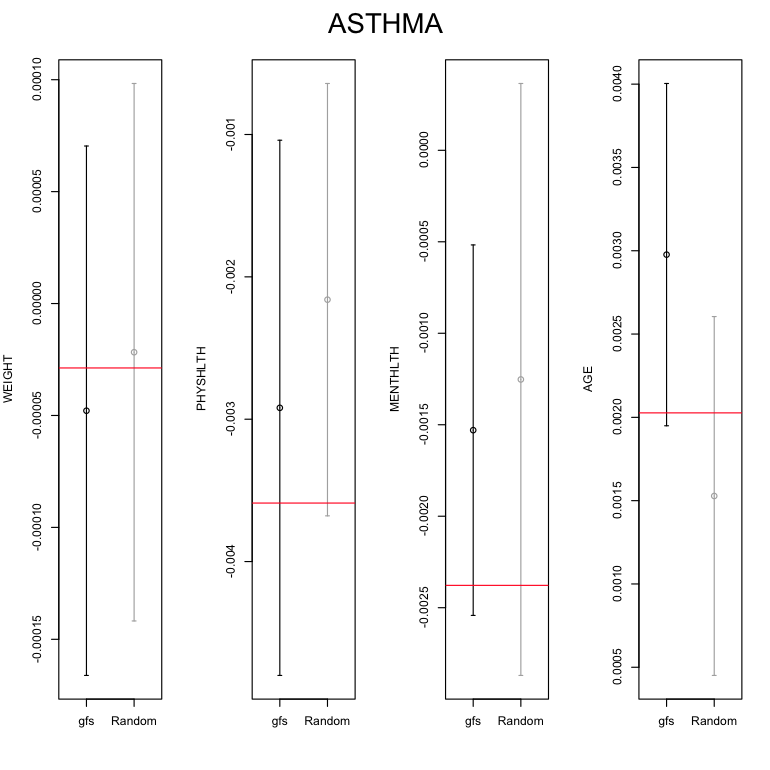}
\caption{\code{GENHLTH}, \code{ALCDAY}, and \code{ASTHMA} under the Logistic model without uniform blocks.}
\label{fig:logistic_plots_A}
\end{figure}

By excluding the blocks containing identical \code{ASTHMA} values we are left with 2216 records, of which 1073 records are single-record blocks and the remaining records are distributed among 263 blocks. Figure \ref{fig:logistic_plots_A} shows the results for the blocks with dissimilar \code{ASTHMA} values. Without the blocks containing identical \code{ASTHMA} values, the bias and the sampling variance of coefficient estimates over the linked data set are reduced compared to when they were included and compared to the random permutations. Nonetheless, the improvement over the random permutations with the logistic model in either case is not as pronounced when compared to the results under the linkage that is generated using the linear regression linkage model. 

\subsubsection{Joint generalized linear linkage models}

\begin{figure}[t]
\centering
\includegraphics[width=0.45\textwidth]{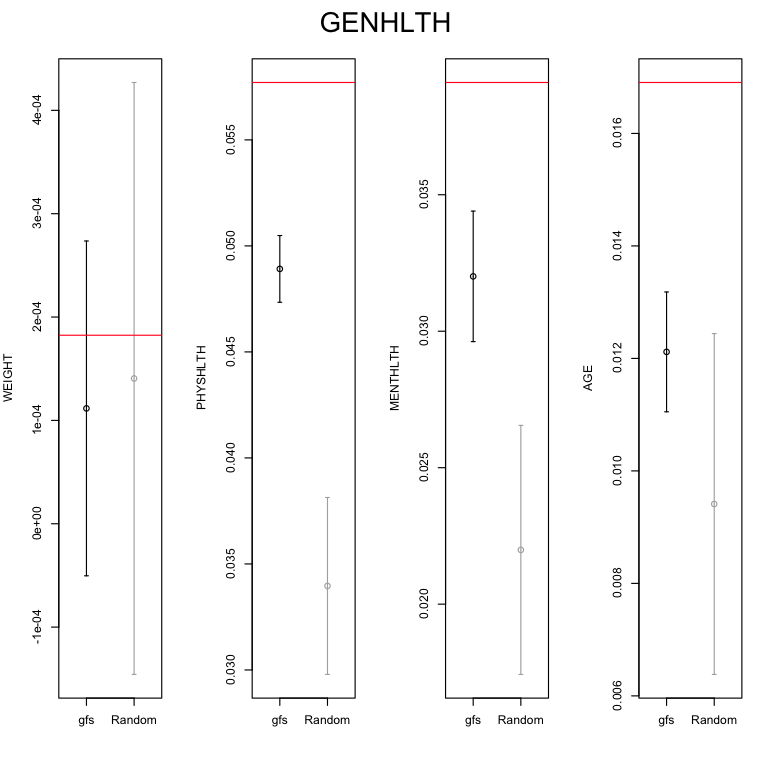} \includegraphics[width=0.45\textwidth]{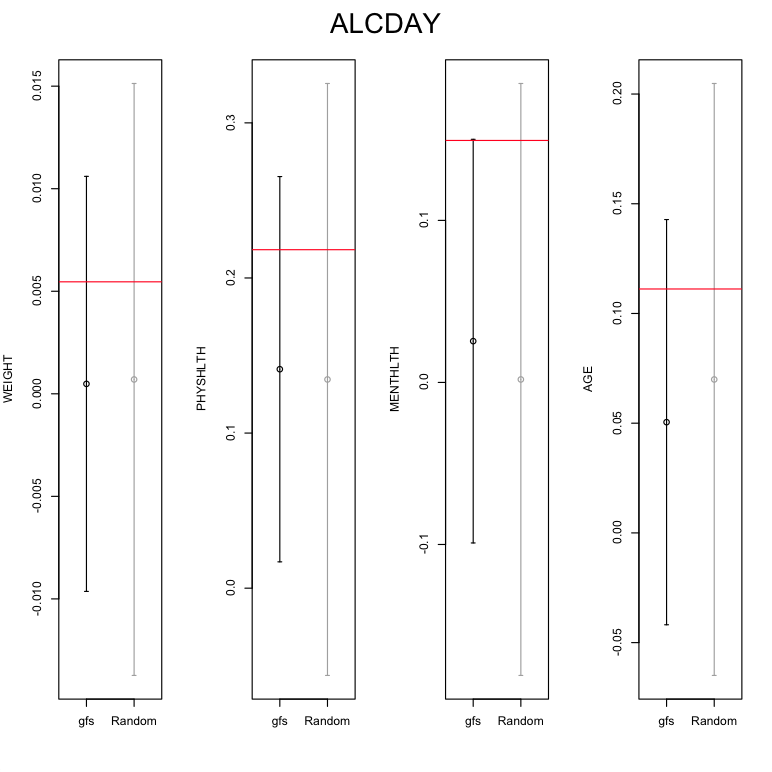}\\
\includegraphics[width=0.45\textwidth]{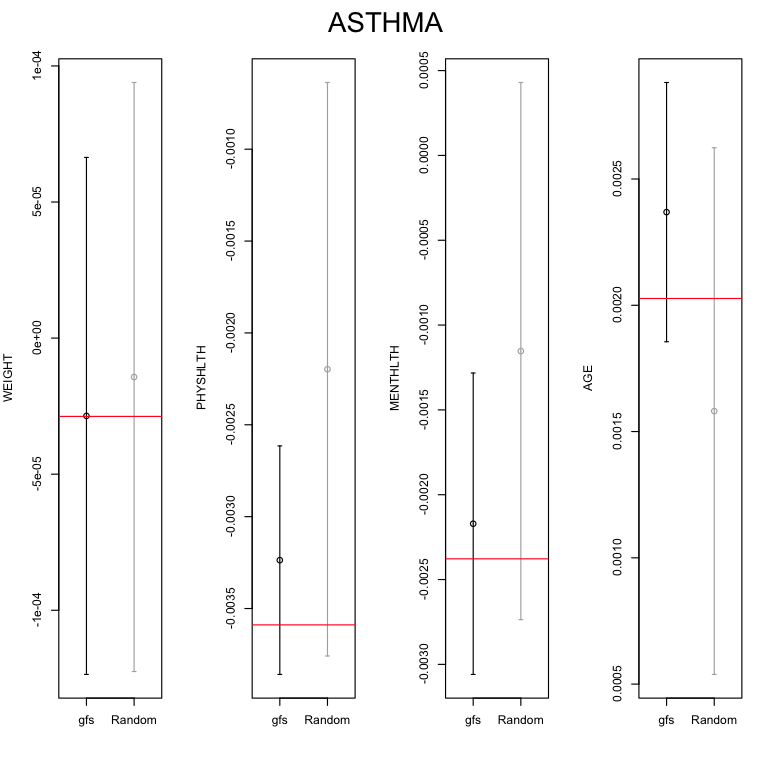}
\caption{\code{GENHLTH}, \code{ALCDAY}, and \code{ASTHMA} under combination of models.}
\label{fig:joint_plots}
\end{figure}

Figure \ref{fig:joint_plots} displays the results for different generalized linear models with one covariate, when permutations are generated using a combination of two models. The joint use of two linkage models results in similar or improved operating characteristics compared to each of the linkage models separately. Specifically, the bias and variability of the interval estimates were similar or smaller than those observed for either the linear regression or the logistic regression linkage models. The improvement over the linear regression model is not substantial in this case, because the associations in the linear regression model are more informative than those under the logistic regression model, as evidenced by our analysis of their individual results.

\subsubsection{Correct matches}

In this simulated example we know the 3,200 true links, so we can compare the number of true links under the proposed method and the number of correct links when rows are permuted at random. Table \ref{tab:all} shows the average number of correctly linked records and the standard deviation of correct links over 10 permutations for each model. 

There is a significant improvement in the number of correct links under the linear regression model and the joint model compared to random permutation sampling. These results mimic the results for the bias and sampling variance of estimates for the slopes of the generalized linear models with a single covariate. Excluding the 1,073 blocks with only one record from each file, the random permutation selection approach yields on average 657 correct links in blocks with more than one record from each file. Using multiple linkage models resulted in 806 correct links in blocks with more than one record from each file. This is an improvement of approximately 23\%.

With the current data set, the number of correctly linked records using random samples is similar to the number of correctly linked records using the logistic regression linkage model. This result is a reminder that the aim of many file linkage applications is not necessarily to correctly link individual records, rather to allow for downstream analyses of the linked data set, and these goals are not necessarily equivalent. In blocks with non-uniform response values, the proposed method with the logistic regression model results in 1,335 correct matches on average, on par with the random permutations method (Table \ref{tab:excl}); however, compared to random permutation sampling, the proposed method using the logistic regression linkage model results in smaller sampling variability and some reduction in the bias of the slopes in the generalized linear models with a single covariate (Figures \ref{fig:logistic_plots} and \ref{fig:logistic_plots_A}).

\begin{table}[H]
\caption{All Blocks}
\begin{tabular}{l|ccc}
    Model & Correct Links & Correct Links excl. singleton blocks & Std. Deviation\\ \hline
    Random & 1730 & 657 & 34 \\
    Normal & 1872 & 799 & 22 \\
    Logistic & 1725 & 652 & 31 \\
    Joint & 1879 & 806 & 27 \\
\end{tabular}
\label{tab:all}
\end{table}

\begin{table}[H]
\caption{Excluding Blocks with Identical values for \code{ASTHMA}}
\begin{tabular}{l|ccc}
    Model & Correct Links & Correct Links excl. singleton blocks & Std. Deviation\\ \hline
    Random & 1338 & 265 & 17 \\
    Logistic & 1335 & 262 & 22 \\
\end{tabular}
\label{tab:excl}
\end{table}

\section{Conclusions}
\label{sc:conclusion}

Probabilistic record linkage methods are becoming increasingly relevant in epidemiology and the social sciences, because they allow researchers to perform complex analyses using granular information that is not available in one data set. The approach presented by \citet{Gutman} and implemented in the \pkg{gfs\_sampler} package mitigates some of the limitations of other approaches to probabilistic record linkage. The \pkg{gfs\_sampler} packages employs a Bayesian record linkage approach that incorporates data that is exclusive to one of the files as well as data common to both files.

We have presented an example to illustrate the improvements that can be achieved by using the proposed probabilistic linkage method. The analysis shows that the choice of model is important in improving the accuracy of the linkage as well as the accuracy and precision of estimates of the parameters of interest. It is important to note that the aim of the method is not necessarily to correctly link individual records, but rather to estimate associations between variables dispersed across two files. As shown in our example, in some cases there is little information from which to
infer which record from one file is related to a record from the other file. If the association between a binary response variable and the explanatory variables is of interest, this is not a concern for inference, because when all the response variables within a certain block are similar, any set of linked records provides identical information; however, if other relationships in the data are of interest, they should be included during the linkage process. Moreover, our analysis shows that including scientifically important correlations, as well as those that are not, improves the linkage and the estimates of parameters of interest. This is similar to the concept of congeniality when implementing multiple imputation \citep{meng1994multiple}. 

In many applications of file linkage, adjustments for possible errors in the linkage are commonly neglected, but such errors arise regardless of which linkage procedure is applied \citep{Shlomo2019}. A number of approaches have been proposed to measure and account for these errors in regression settings \citep{Shlomo2019}. The proposed Bayesian approach enables the estimation of linkage errors by sampling from the posterior distribution of possible linkage permutations. This approach is not restricted to regression models, and it allows for propagation of linkage errors to estimate any statistic after the linkage is performed. 

The implemented approach that is based on multiple imputation is computationally efficient. First, its underlying implementation is in \proglang{Python}, which enables efficient MCMC sampling. Second, the linkage structure is saved in an efficient data structure. Specifically, we only save permutations of indices so that the original files are not saved for every sampled linkage. Third, because in many applications, record linkage serves as a tool to investigate specific scientific questions, our program produces multiple linkage structures that can be used to estimate any parameters of interest. The more computationally expensive process of linkage can be performed once, after which researchers can perform multiple analyses and combine their results using common multiple imputations rules. Fourth, the method allows for unbalanced blocks in the files to be linked by relying on a monotone missing data pattern and imputing mismatched records in only one of the files as a part of the sampling process.  

\begin{comment}
Since calculating statistics on the full data set with our approach relies on multiple imputation, this bias correction would be applied to each imputation instance. A different approach proposed by \citet{Goldstein} attempts to resolve contested matches by treating them as missing data. A common issue in all such solutions is that the truth is not known, even for a subset of records. Nonetheless, it is worth considering how linkages can be better understood or improved in conjunction with the linkage algorithm itself---including the algorithm and implementation we have presented in this article. 

One possible limitation of the linking algorithm is that sampling from the distribution of linkage permutation can be relatively slow.
\end{comment}

We have used blocking to increase efficiency and scalability; however, in cases when the blocking variables are recorded with errors, blocking may exclude true links and influence subsequent inferences. An area of improvement for the current method is to extend the sampling algorithm to allow for errors in blocking variables \citep{dalzell2018regression}. While this could mitigate the potential of erroneous blocking to skew inferences, such an addition could increase the computational complexity of the sampling algorithm to the point where it would become computationally prohibitive for large data sets. Another area of improvement for the current implementation is support for additional regression models. Hierarchical regression models and two-part models, such as zero inflated models, are potential additions. The design of the code allows for modular addition of new distributions. Lastly, the use of parallel computing and targeted improvements to the MCMC sampling procedure akin to those proposed by \citet{zanella2020informed} may improve the performance of the algorithm on large data sets. Selection of models that should be used in the linkage process is another area for further inquiry and improvement. The best choice of response and explanatory variables are variables that are highly correlated; however, this correlation cannot be computed across the separate files. External data sources could inform selection, but incorporating this information in the modeling process is an area of further research. In addition, including more models has the potential to improve the linkage performance, but it may also increase computational complexity. Defining a way to measure the contributions of additional variables is another area for future research. 

In conclusion, we describe a computationally efficient algorithm to perform file linkage with variables that are exclusive to one file or are recorded with errors in one of the files. The algorithm generates multiple linkage structures, allowing for the propagation of errors in linkage through multiple imputation. In addition, we illustrate the use of the \pkg{gfs\_sampler} record linkage package on a real data set. This example can serve as a starting point for researchers interested in implementing a Bayesian procedure to link records across files in the absence of unique identifiers.

\section*{Acknowledgements}
This research was partly supported through a Patient-Centered Outcomes Research Institute (PCORI) Award ME-1403-12104. Disclaimer: All statements in this report, including its findings and conclusions, are solely those of the authors and do not necessarily represent the views of the PCORI, its Board of Governors or Methodology Committee.

We would like to thank Preston Schwartz, for his early contributions to development and testing of the \pkg{gfs\_sampler} package.
\bibliography{refs}
\newpage

\section{Appendix}
\label{sc:appendix}
\subsection{Installation and Getting Started Instructions}

\subsubsection{Python package}

The \pkg{gfs\_sampler} is available on the PyPI repository: \url{https://pypi.org/project/gfs\_sampler/}.

The \pkg{gfs\_sampler} is supported in \proglang{Python} versions 3.6 and 3.7 because our code relies on the \pkg{pymc3} and \pkg{theano} packages. These two packages are not supported in \proglang{Python} versions  3.8 and newer. We recommend using Anaconda for managing virtual environments and package versions in \proglang{Python}. The following command will create an environment named \code{<env_name>} with the required packages for the \pkg{gfs\_sampler} package:
\begin{knitrout}
\definecolor{shadecolor}{rgb}{0, 0, 0}\color{shcolor}\begin{kframe}
\begin{verbatim}
> conda create -n <env name> numpy pandas pymc3==3.6 
+ 	mkl mkl-service theano==1.0.4
\end{verbatim}
\end{kframe}
\end{knitrout}

Once the environment is created, it can be activated and the \pkg{gfs\_sampler} package can be installed to the virtual environment with the following command using \code{pip}: 
\begin{knitrout}
\definecolor{shadecolor}{rgb}{0, 0, 0}\color{shcolor}\begin{kframe}
\begin{verbatim}
> source activate <env_name>
> pip install gfs_sampler
\end{verbatim}
\end{kframe}
\end{knitrout}

The \pkg{pymc3} sampling process produces warnings and logs that one may want to remove. Using the \pkg{warnings} and \pkg{logging} modules, these warnings and logs can be removed:
\begin{knitrout}
\definecolor{shadecolor}{rgb}{0.969, 0.969, 0.969}\color{fgcolor}\begin{kframe}
\noindent
\ttfamily
\hlstd{}\hlkwa{import\ }\hlstd{warnings}\hspace*{\fill}\\
\hlstd{}\hspace*{\fill}\\
\hlstd{warnings}\hlopt{.}\hlstd{}\hlkwd{simplefilter}\hlstd{}\hlopt{(}\hlstd{}\hlstr{"ignore"}\hlstd{}\hlopt{,\ }\hlstd{}\hlkwc{UserWarning}\hlstd{}\hlopt{)}\hspace*{\fill}\\
\hlstd{}\hspace*{\fill}\\
\hlstd{}\hlkwa{import\ }\hlstd{logging}\hspace*{\fill}\\
\hlstd{}\hspace*{\fill}\\
\hlstd{}\hlslc{\#\ these\ commands\ should\ be\ just\ before\ running\ gfs.sample()\ command}\hspace*{\fill}\\
\hlstd{logger\ }\hlopt{=\ }\hlstd{logging}\hlopt{.}\hlstd{}\hlkwd{getLogger}\hlstd{}\hlopt{(}\hlstd{}\hlstr{"pymc3"}\hlstd{}\hlopt{)}\hspace*{\fill}\\
\hlstd{logger}\hlopt{.}\hlstd{}\hlkwd{setLevel}\hlstd{}\hlopt{(}\hlstd{logging}\hlopt{.}\hlstd{ERROR}\hlopt{)}\hlstd{}\hspace*{\fill}
\mbox{}
\normalfont
\end{kframe}
\end{knitrout}

\subsubsection{R package}

1. Installing GFS: With \code{devtools}, use \code{install\_github} to fetch and install the package from \url{https://github.com/edwinfarley/GFS}:

\begin{knitrout}
\definecolor{shadecolor}{rgb}{0.969, 0.969, 0.969}\color{fgcolor}\begin{kframe}
\begin{alltt}
\hlstd{> }\hlkwd{library}\hlstd{(}\hlstr{"devtools"}\hlstd{)}
\hlstd{> }\hlkwd{install_github}\hlstd{(}\hlstr{"https://github.com/edwinfarley/GFS"}\hlstd{)}
\hlstd{> }\hlkwd{library}\hlstd{(}\hlstr{"GFS"}\hlstd{)}
\end{alltt}
\end{kframe}
\end{knitrout}

Alternatively, clone the repository yourself from the same link and then use \code{install\_local}, passing the path to the local directory that contains the package. If you only intend to use the R package for analysis, then no further steps are required; however, it is also possible to run the sampling method directly from \proglang{R} and the remaining steps explain the process necessary to use this functionality.

2. Set up the \proglang{Python} environment: We recommend using an Anaconda distribution of \proglang{Python} 3.6 and 3.7 for managing environments. Use the included \code{create\_python\_environment()} function to set up a Conda environment with all the required packages. This function takes a single argument, \code{conda\_envname}, the name of the environment to be created. The new environment will include the following packages: numpy, pandas, pymc3, mkl, theano, and mkl-service.

3. Installing \proglang{Python} components: The \proglang{Python} components will be installed automatically when the \code{permute\_inputs} is called if no \proglang{Python} directory is found; however, the package does include a \code{py\_setup()} function in R. This function call clones the repository at \\ https://github.com/edwinfarley/GFSPython to a directory named ``Python'' in the GFS package directory.

4. Ready to go: These are the required steps for preparing to run the \code{permute\_inputs} function to sample permutations. Take a look at the documentation with \code{?permute\_inputs} in R to see information about the arguments and sampling parameters. Be sure to pass the name of your Conda environment to the \code{conda\_env} parameter when using \code{permute\_inputs}. If the \proglang{Python} components have not been installed manually with the provided function, they will be installed automatically when \code{?permute\_inputs} is run for the first time.

\end{document}